\documentclass[letterpaper,aps,longbibliography,notitlepage,nofootinbib]{revtex4-2}
\usepackage[latin9]{inputenc}
\usepackage[]{mathbbol}
\usepackage{amsmath}
\usepackage{amssymb}
\usepackage[scr=boondoxo,cal=zapfc]{mathalfa}
\usepackage{graphicx}
\usepackage[hidelinks]{hyperref}
\usepackage{setspace}
\usepackage{wrapfig}
\usepackage{enumitem}
\usepackage[overload]{textcase} 
\usepackage{footnotehyper}

\DeclareSymbolFontAlphabet{\mathbb}{AMSb}
\DeclareSymbolFontAlphabet{\mathbbl}{bbold}

\DeclareSymbolFont{extraup}{U}{jkpsyc}{m}{n} 
\DeclareMathSymbol{\vardiamondsuit}{\mathalpha}{extraup}{113} 

\makeatletter

\pdfpageheight\paperheight
\pdfpagewidth\paperwidth
\renewcommand\frontmatter@abstractwidth{.787\textwidth} 

\pdfoutput=1

\def\@fnsymbol#1{\ensuremath{\ifcase#1\or *\or \dagger\or \ddagger\or
   \mathsection\or \mathparagraph\or \|\or \#\or \vardiamondsuit \else\@ctrerr\fi}}

\makeatother

\begin{document}

\title{Phenomenology of Holography via Quantum Coherence on Causal Horizons}

\author{\vspace{2pt}Ohkyung Kwon\footnote[0]{{\small E-mail: \href{mailto:kwon@uchicago.edu}{\nolinkurl{kwon@uchicago.edu}}}}} 
\affiliation{\vspace{2pt}University of Chicago}

\begin{abstract}

\linepenalty=310

\vspace*{-6.25pt}\vspace*{4pt}There is much recent development towards interferometric measurements of holographic quantum uncertainties in an emergent background space-time. Despite increasing promise for the target detection regime of Planckian strain power spectral density, the foundational insights of the motivating theories have not been connected to a phenomenological model of observables measured in a realistic experiment. This work proposes a candidate model, based on the central hypothesis that all horizons are universal boundaries of coherent quantum information\,--- where the decoherence of space-time happens for the observer. The prediction is inspired by 't~Hooft's algebra for black hole information that gives coherent states on horizons, whose spatial correlations were shown by Verlinde and Zurek to also appear on holographic fluctuations of causal boundaries in flat space-time (conformal Killing horizons). Time-domain correlations are projected from Planckian jitters whose coherence scales match causal diamonds, motivated by Banks' framework for the emergence of space-time and locality. The universality of this coherence on causal horizons compels a multimodal research program probing concordant signatures: An analysis of cosmological data to probe primordial correlations, motivated by Hogan's interpretation of well-known CMB anomalies as coherent fluctuations on the inflationary horizon, and upcoming 3D interferometers to probe causal diamonds in flat space-time. Candidate interferometer geometries are presented, with a modeled frequency spectrum for each design. 

\end{abstract}
\maketitle

\setstretch{1.09}

\section*{\NoCaseChange{Introduction}}

The Fermilab Holometer\,\cite{chou2016,chou2017,chou2017a,Richardson_2021}, proposed in 2009 and in operation through 2019, was the first instrument commissioned to directly search for quantum uncertainties in space-time, conjectured to be measurable in the regime of Planckian strain power spectral density if the background space-time is emergent from a quantum system with holographic total degrees of freedom (``Holographic Noise'')\,\cite{Kwon:2014,Hogan_2017,Hogan:2015b,Hogan:2016}. This line of research, initially based on Hogan's conceptual insights in 2007 about the large correlations necessary for such dimensional reduction of information in the Planckian degrees of freedom of background space-time\,\cite{Hogan:2008a}, was met with controversy. In recent years, however, the motivating theoretical ideas have been reproduced with widely accepted mathematical methods by Banks, Verlinde, and Zurek~\cite{Verlinde_2021,Verlinde:2019ade,Banks_2021,Verlinde:2022hhs,banks2023fluc}. The current state of foundational insights has yet to result in a phenomenological model of observables measurable in a real experiment,\footnotemark[2]\;which is the aim of this manuscript. Presently, two major experiments are being advanced to succeed the Holometer with significant new technologies: the GQuEST experiment being started by a Caltech-Fermilab team\,\cite{GQuESTPublic,McCuller:2022hum,vermeulen2024}, and an experiment already under commissioning since 2019 at Cardiff University\,\cite{grote2020,QIPublic} as part of the Quantum-enhanced Interferometry for New Physics~(QI) research program under the UK's Quantum Technologies for Fundamental Physics (QTFP) initiative. The QI consortium experiment at Cardiff was specifically designed with the future capability to construct the most promising configurations proposed here.

A standard perspective, based on the framework of local effective field theory, is that quantum effects in space-time manifest at the Planck scale $\ell_P \equiv c\,t_P$, where $t_P\equiv \sqrt{\hbar\hspace{.07em} G/ c^5} = 5.4\times 10^{-44}\,$sec. However, there are strong indications that space-time is emergent from the thermodynamics of covariant causal structures\,\cite{Jacobson1995,Jacobson:2015hqa}, and that the holographic entropy of a black hole $S_{\mathrm{BH}} = k_B A \,/\, 4\hspace{.07em}\ell_P^2$ (where $A$ is the area of the horizon) can be generalized to a covariant entropy bound that might apply universally\,\cite{tHooft1993,Susskind1995,Bousso:1999xy}. If one takes this seriously in flat space-time, the density of information in a causal diamond\,--- the number of microscopic degrees of freedom per volume\,--- decreases as the inverse of the system scale. This picture necessitates that the space-time degrees of freedom have large correlations in the \textit{infrared}, holistically dependent on the extent of the system, nonlocally encoded via the bounding causal structures.

\enlargethispage{7.5pt}
An estimate of holographic space-time uncertainties in this picture\,\cite{Kwon:2014,Hogan_2017,Hogan:2015b,Hogan:2016,Verlinde_2021,Verlinde:2019ade,Banks_2021,Verlinde:2022hhs,banks2023fluc} says that for an observational system of scale $L= c\hspace{.07em}\tau$, a causal diamond delimiting its measurement should embody fluctuations of variance $\langle (\Delta L\hspace{.07em}/L)^2\hspace{.07em}\rangle \sim t_P/\tau$. As a frequency-domain signal, the strain $h\equiv\Delta L\hspace{.07em}/L$ has a power spectral density (PSD) of \:\!$\sim\:\!\!t_P$ (in units of inverse frequency) over a bandwidth \:\!$\sim\:\!\!1/\tau$ (in units of frequency). This ``Planckian random walk'' scaling is analogous to the standard quantum positional uncertainty $\langle\hspace{.07em}\Delta x^2\hspace{.07em}\rangle > \hbar\hspace{.07em}\tau/m$ of a mass~$m$ over time~$\tau$, and corresponds to the scale of quadrupolar distortions a black hole horizon needs in order to radiate at the standard Hawking flux, one graviton of wavelength~$c\hspace{.07em}\tau$ per time~$\tau$. Heuristically, if each causal diamond smaller than the system size adds a Planck-scale jitter with a coherence scale matching its boundary, they accumulate like a random walk over \:\!$\sim\:\!\!\tau/t_P$ nested layers, creating spacelike fluctuations much larger than expected in standard local effective field theory by the same factor.\hspace{4pt} 

This paper presents a candidate model of quantum space-time phenomenology in this regime of Planckian strain power spectral density. A few important theoretical developments are reviewed, moving beyond the conceptual arguments above and rigorously establishing that low-energy, macroscopic quantum space-time could have significant departures from classicality of this magnitude due to holography. A set of general principles is proposed to encapsulate the characteristic physics implied by these developments. Two geometric configurations are identified in which real-world observables measured in an interferometer are expected to couple to the conjectured effect, and their design parameters optimized such that the magnitude of the quantum space-time uncertainty and the instrumental response are set at levels attainable in realistic state-of-the-art instruments in the near future. Experimental signatures are modeled, including time-domain and frequency-domain spectra for the geometries proposed. 

The reader is strongly encouraged to give careful consideration to the key literature cited in the following section. These seminal works represent important, fundamental departures from the mainstream paradigms of the quantum gravity community.\vspace{-6.5pt}

\footnotetext[2]{A prediction by Zurek and colleagues has been released since a preprint of this manuscript was posted. However, the present author finds that this prediction\,\cite{Zurek_2022,Zurek:2022xzl,LiPRD} sacrifices some of the most important foundational insights from the currently available body of theoretical work in order to adopt a more precisely defined calculational framework.\vspace{2.25pt}}

\section*{\NoCaseChange{Background and Motivation}}

It has long been understood that standard local effective field theory results in infrared catastrophes at surprisingly small scales. The Cohen-Kaplan-Nelson bound\,\cite{CohenKaplanNelson1999}, derived by simply counting all the possible modes of the QFT in a given system volume and imposing the black hole entropy bound while insisting that there be no states with a Schwarzschild radius larger than the system size, shows that for an ultraviolet cutoff at the QCD scale $\Lambda\sim 200\,\mathrm{MeV}\approx1.7\times 10^{-20}\,m_P$, the system is subject to an infrared cutoff at the $L\sim\Lambda^{-2}\sim4\times10^{-39}\,\ell_P\approx60\,$km scale (in Planck units) to avoid such failures. This UV-IR scaling coincides neatly with the Planckian strain spectral density scaling discussed here: if the system size $L$ serves as an IR bound, the corresponding Planckian random walk uncertainty for the underlying space-time $\langle\hspace{.07em} \Delta L^2\hspace{.07em}\rangle^{1/2} \propto \sqrt L \sim \Lambda^{-1}$ matches the localization scale for field modes at the UV cutoff~\cite{Hogan:2014zma}.\hspace{3.5pt}

This problem has not attracted much serious effort towards a solution, despite being one of the most significant failures framing the fundamental issues in quantum gravity, partly because of blind spots in our standard frameworks, and partly because addressing these shortcomings requires difficult sacrifices of key notions we have presupposed as essential building blocks in our physical theories. The lesson of the CKN bound is that a definite background space-time manifold that is smooth all the way down to Planckian microstates simply contains too much information (relative to no background), when the true degrees of freedom of the quantum space-time system are far more limited, necessitating large correlations in the background space-time itself. Banks has argued for years that the AdS/CFT duality is failing to account for these background degrees of freedom in its holographic projection, and that the cosmological constant\,--- ``the worst prediction in physics'' at 122 orders of magnitude smaller than QFT-based estimates\,--- is a boundary condition for these omitted degrees of freedom~\cite{banks2023holography,Banks:2018jqo,Banks:2018aed,Banks:2020dus,*banks2022old,banks2023hilbert,*banks2023impossibility,*BanksGRF2023}. In AdS space, the built-in boundary with its negative curvature structure hides this infrared issue for most quantum fields by construction. In contrast, consider the problems encountered trying to describe quantum space-time modes using gravitational effective field theory in a finite bounded volume, say, a causal diamond\,\cite{KITPdiscussion}: we want to use unitarity, which means we have to use a non-Lorentz-invariant gauge, and furthermore, the only gauge-invariant things we calculate in gravity are scattering amplitudes at infinity\,--- if we insist that the time slices stay within the causal boundary, as one ought for the problem discussed here, the solution would require a non-translation-invariant gauge. There is no known well-defined finite space-time setup for gravitational effective field theory that meets these conditions.\footnote[7]{This is why the standard effective field theory calculation for quantum gravitational fluctuations in an interferometric measurement\,\cite{Kwon:2014,carney2024} should not be considered valid for interrogating the kind of foundational issues considered here.}

The same flaw was alternatively explained from the general relativist's perspective in an illuminating essay by Hollands and Wald~\cite{Hollands_2004}. The key issue is that standard quantum field theory does not fully capture the nonlocality of quantum mechanics even in the low-energy effective degrees of freedom. In quantum mechanics, nothing happens on definite paths in space-time (world lines); everything exists in a superposition state. The active gravity from the mass-energy of such a state is necessarily nonlocal, and so is its effect on space-time, including the distortion of causal structures. This concern is usually dismissed as an insignificant correction, and only considered in the context of thought experiments where extremely large coherent quantum states of matter can actually create indeterminate superpositions of different causal orderings\,\cite{Zych_2019}. However, the Hollands-Wald manuscript demonstrates that these nonlocal correlations in the background metric in fact entirely invalidate the basic underlying construction of QFT in the context of the space-time vacuum. The renormalization of a field theory requires subtracting infinities that are interpreted as the physical modes of the vacuum, but this process is necessarily global in character: ``An individual mode will have no way of knowing whether its own subtraction is correct unless it `knows' how the subtractions are being done for all the other modes''\,\cite{Hollands_2004}. To enforce locality and covariance, the standard QFT prescription for decomposing a quantum field into modes resorts to acausally and unphysically constraining the system at infinity independently in all directions. This is true even in a low-energy, flat space-time example, where the renormalization prescription subtracts the vacuum energy of the modes that hypothetically would have been present in a globally Minkowskian universe, instead of the ones that are actually present in a particular QFT system. In the real universe, distortions in causal structure caused by modes with wavenumber $k\rightarrow0$, including events at distance $R\rightarrow\pm\infty$, reshape the QFT mode decomposition and their mapping onto events in space-time on smaller scales. The typically assumed classical background with definite causal structure, on which the fields are defined, is subtly altered at all scales in a way that makes it impossible to achieve local and covariant renormalization without considering the holistic aspects of the QFT system. According to Hollands and Wald, this failure to correctly subtract from the low energy modes of the QFT is the source of the cosmological constant problem. A meticulous, informative manuscript by Stamp\,\cite{Stamp_2015} has also laid out similar fundamental issues, with a proposed solution of rewriting the path integral formalism of QFT with world lines that are reformulated to include these nonlocal correlations in space-time.

A candidate framework for constructing locality from a quantum system is found in the Holographic Space-Time (HST) theory proposed by Banks and Fischler\,\cite{Banks2011,Banks:2018aed,Banks_2020,Banks:2020dus,*banks2022old,banks2023hilbert,*banks2023impossibility,*BanksGRF2023}. In this theory, which is formulated without any background space-time, every causal diamond has its own Hilbert space, whose dimension scales holographically with the size of the causal boundary. Nested causal diamonds are subspaces via tensor factorizations, and overlapping causal diamonds are similarly handled via tensor products. Information is delocalized within the boundaries of causal diamonds, which determine the entanglement of geometrical states. Entangled states of different world lines undergo ``fast scrambling'' outside of intervals set by causal diamonds that scale with their separation. The viewpoint is that ``The Holographic Principle tells us that both the causal structure and conformal factor of the space-time geometry are encoded as properties of the quantum operator algebra. The space-time geometry is not a fluctuating quantum variable, but, in general, only a thermodynamic\,/\,hydrodynamic property of the quantum theory.'' This theory is general enough to accommodate field theory in the appropriate limits, although a formulation recovering the Standard Model has not yet been enunciated. 
Importantly, it provides a precisely controlled framework for the amount of correlation between the quantum geometric information contents of different causal volumes demarcated by their respective causal diamonds. Even though a well-defined connection to real space-time observables remains unclear, the framework provides basic principles about the scaling and localization of quantum geometric information that will guide our estimates of time-domain correlations for experimental measurements in the next section. We will rely extensively on the notion that fast scrambling of information happens on the boundaries of causal diamonds (conformal Killing horizons), and that we may measure coherent quantum effects of geometry in our time-domain correlations when our observables are appropriately coupled to space-time within those causal boundaries.

\enlargethispage{2.5pt}
For the nonlocal spatial correlations, it is instructive to take a brief detour to consider a concrete example that cannot be captured by a field theoretic approach\,\cite{krishogan}. Consider a case where massive particles are decaying to EPR-like states of oppositely propagating photons, and the direction of decay is indeterminate\,--- as in, the photons are created in $S$-wave states, as isotropic superpositions of all directions. The gravitational shockwaves\,\cite{Aichelburg1971,DRAY1985} from such decays result in coherent quantum gravitational fluctuations of causal diamond boundaries on all scales. Statistically, the angular correlations are \textit{independent of scale} and approximately quadrupolar. Due to the permanent gravitational memory effect, this can accumulate. For a measurement system in which the number of decaying particles is limited by the gravitational binding energy corresponding to its size, with an ultraviolet cutoff at the Planck mass, the amount of distortion in causal structure scales linearly with the duration of the measurement, just as in the holographic Planckian random walk scaling described above (a generalization to a null perfect fluid was done in Ref.\,\cite{mackewicz2024}). This example is not intended to be an appropriate model for our proposed experiment, where the interferometer measures fluctuations in empty space-time and there is no active source at the center of the causal diamond. In a realistic model, the fabric of space-time is constructed without a background as in the HST theory, via intersecting causal diamonds whose vacua are filled with virtual states that pop in and out of existence (zero mean, nonzero mean square power). The coherent quantum gravitational fluctuations on these causal diamond boundaries originate from the relationships among different observers and the incoming\,/\,outgoing information among their respective causal diamonds.

A pathbreaking model of quantum information on black hole horizons derived by 't~Hooft\,\cite{Hooft:2016itl,*Hooft:2016cpw,*Hooft2016_,Betzios_2016,tHooft2018,Hooft:2018gtw,*Hooft:2018syc,*hooft2019,Gaddam_2022}\footnote[4]{These works comprise arguments well grounded in classical quantum mechanics and general relativity, and should be regarded separately from the more controversial expositions on interpretations of quantum mechanics by the same author.} helps us develop intuition for what kind of coherent quantum gravitational fluctuations are possible on causal horizons when quantum modes are traveling in and out of the boundary. This model separates the transverse operators from the radial ones, and combines a quantum treatment of the radial components of particle states together with the radial components of gravitational frame dragging. For a particle state entering the horizon of an eternal Schwarzschild black hole, by correctly accounting for the gravitational back reaction across the surface of the horizon, the model finds it entangled with another particle state at its antipode leaving the horizon. The antipodal entanglement is in both space and time\,--- between antipodes in 3-space, with a sign inversion in time\,--- such that the position operator of an outgoing particle is associated with the momentum operator of an incoming one at the antipode, and vice versa. The resulting topology appears quite strange, but this solution is mathematically shown to uniquely restore unitarity to black hole information when the effects of gravitational back reaction are properly included. \textit{All of the information is nonlocally encoded on the boundary, as coherent states on the horizon.} Reminiscent of the Banks-Fischler theory where causal diamond boundaries are fast scramblers, the causal horizon in this black hole model constitutes a kind of quantum-classical boundary\,--- a boundary of coherent quantum information. In this perspective, to an observer outside the boundary, the delocalized interior may as well not exist. The entire black hole is a single quantum object, akin to a hydrogen atom, and its quantum states can be expanded in term of spherical harmonics. Unlike the previous example with approximately quadrupolar correlations, the expansion here only accepts odd harmonics due to the antipodal antisymmetry between incoming and outgoing modes. The algebra of the basis operators, which are written in null coordinates $u^\pm$, $p^\pm$ and describe delocalized states that extend across the entire horizon, is:\vspace{-.25pt}
\begin{subequations}
\begin{align}
		&u^\pm(\Omega)=\sum\limits_{\ell,m}u^\pm_{\ell m}Y^{}_{\ell m}(\Omega)&& \quad\ 
		p^\pm(\Omega) =\sum\limits_{\ell,m}p^\pm_{\ell m}Y^{}_{\ell m}(\Omega)&& \quad\ 
		[u^\pm(\Omega),\,p^\mp(\Omega')]=i\delta^2(\Omega,\,\Omega') \qquad \label{1a}\\  
		&[u^\pm_{\ell m},\,p^\mp_{\ell' m'}] =i\delta_{\ell \ell'}\delta_{mm'}&& \quad\ 
		[u^\pm_{\ell m},\,p^\pm_{\ell' m'}] =0\vphantom{\smash[t]{\sum}} && \label{1b}\\ 
		&u^-_\mathrm{out}=\frac{8\pi G} {\ell^2+\ell+1}p^-_\mathrm{in}&& \quad\ 
		u^+_\mathrm{in}=-\frac{8\pi G} {\ell^2+\ell+1}p^+_\mathrm{out}&& \quad\ 
		[u^+_{\ell m},\,u^-_{\ell' m'}]=i\frac{8\pi G} {\ell^2+\ell+1} \delta_{\ell \ell'}\delta_{mm'}\ \label{1c}		
\end{align}\label{hooftcorr}
\end{subequations}\vspace{-6pt} 

A surprising connection was shown by Verlinde and Zurek\,\cite{Verlinde_2021} when a similar angular correlation function was calculated for quantum gravitational fluctuations on causal diamonds (conformal Killing horizons) in holographic flat space-time. A key conjecture is established, that there is a universal effect of metric fluctuations near horizons, because standard thermodynamics near a black hole horizon equally applies near null surfaces associated with boundaries of finite regions in flat space-time, e.g. a Rindler type horizon. Using a topological black hole coordinate transformation on a causal diamond in flat space-time, the paper gives the following correlation function on the bounding 2-sphere:\vspace{-.5pt}
\begin{equation}\label{zurekcorr}
\Big\langle \Delta u^-(\mathbf{r}_1)\, \Delta u^+(\mathbf{r}_2) \Big\rangle = \frac{1}{\sqrt{2\pi}}\, \ell_P L \,\cdot\, \mathbf{G}(\mathbf{r}_1,\, \mathbf{r}_2) \qquad
\mathrm{where} \qquad
\mathbf{G}(\mathbf{r}_1,\, \mathbf{r}_2) = \sum_{\ell,m} \frac{Y^{}_{\ell,m} (\mathbf{r}_1)\, Y^*_{\ell,m} (\mathbf{r}_2) }{\ell^2 + \ell + 1} \ \ \vspace{-3.75pt}
\end{equation} 
Importantly, this correlation scales linearly with the system size $L$ like a Planckian random walk, larger than the one generated by Eq.\,(\ref{hooftcorr}) by a factor $L/(\sqrt{8\pi}\ell_P)$. The smaller incorrect scaling arises from a shortcoming of the 't~Hooft prescription in handling holography: it only quantizes and calculates gravitational effects for the radial parts of the states, and in the transverse direction, just applies a mode cutoff (or angular filter) at the Planck scale where the transverse gravitational effects become significant. This provisionally gives the correct number of degrees of freedom, but a proper quantum treatment of the transverse parts is clearly needed. The Verlinde-Zurek model similarly only calculates correlations for the radial components, but the effect is derived based on a 3D holographic system, borrowing from Marolf's estimation of the quantum width of a black hole horizon from its thermodynamic fluctuations\,\cite{Marolf:2003bb}. 

\enlargethispage{10.5pt}
Verlinde and Zurek have reproduced this outcome in the context of AdS/CFT\,\cite{Verlinde:2019ade}, using thermodynamic arguments that carefully avoid the flaws of field theory pointed out in the Banks-Fischler and Hollands-Wald papers and take into account some of the holistic aspects of the QFT system. This was followed up by an elegant paper by Banks and Zurek\,\cite{Banks_2021}, which generalizes a key result from the AdS/CFT calculation: thermodynamic fluctuations in the modular Hamiltonian of a causal diamond are equal to the entanglement entropy. This provides strong general evidence for the random walk scaling with the size of the holographic quantum system (although the theory retains the limitations in quantizing the transverse directions with effective field theory\,\cite{Banks_2021,banks2023hilbert,*banks2023impossibility,*BanksGRF2023}). The new theory, deriving metric fluctuations from near-horizon thermodynamics via a conformal field theory of the vacuum states, applies broadly to flat space, the cosmological horizon of dS, and AdS Ryu-Takayanagi diamonds, but not to large diamonds or black holes in the bulk of AdS. This general result and the associated correlation functions were shown in a later Verlinde-Zurek model to originate from the gravitational shockwaves of vacuum fluctuations in the causal diamond\,\cite{Verlinde:2022hhs}, consistent with the toy model interpretation of Refs.\,\cite{krishogan,mackewicz2024}, although the localized quantization on near-horizon planar light sheets adopted here results in a non-scale invariant angular scaling at high $\ell$ (more later). Banks and Fischler have also proposed another approach connecting the entropy fluctuations to the 't~Hooft shockwave commutation relations~\cite{banks2023fluc,banks2023hilbert,*banks2023impossibility,*BanksGRF2023}.

It should be noted that all of these correlations derived using framework of effective field theory for better calculational control are conceptually different from the 't~Hooft models. They do not retain the insight from 't~Hooft that the space-time correlations are described by coherent quantum states that macroscopically extend across the entire horizon, \textit{which cannot be described by local metric fluctuations}. Indeed, while the Verlinde-Zurek calculation gives the same spherical harmonic modes that respect the spherical symmetry of the system, the quantum operators are mapped onto definite metric fluctuations, and the correlations described by classical Green's functions instead of the $S$-matrix used in 't~Hooft's ``hydrogen atom'' model of the quantum black hole to handle information propagating in and out of the system. As such, the Verlinde-Zurek model also does not feature any antipodal antisymmetry between entangled incoming and outgoing modes like the 't~Hooft model does, and does not select for the odd harmonics.

A most compelling reason to study coherent quantum fluctuations on horizons comes from signatures of primordial fluctuations\,--- the only known observations of active gravitational effects from quantum states. It is generally understood that the large scale structure we see today originates from distortions in space-time caused by quantum fluctuations in the early universe, which became permanently ``frozen in'' as they crossed the inflationary horizon. The cosmic microwave background on large angular scales also preserves a mostly intact map of the gravitational potential on our cosmic horizon during this process. In the absence of a theory of quantum gravity, existing models of inflation resort to patching QFT modes onto a classical background space-time. This works very well at smaller angular scales $\ell\gtrsim30$, where most of the analysis is done to avoid cosmic variance, or the difference between possible realizations of the universe from a quantum process. However, at the largest angular scales $\ell\lesssim30$, where the primordial effects are expected to be dominant, numerous well-known anomalies persist with $p$-values in the 0.1\,--\,0.5\% range\,\cite{Muir_2018,jones2023universe,Copi_2009,*Schwarz_2016}, which Hogan, Meyer, and colleagues have shown only become more compelling with more careful analysis\,\cite{Hagimoto_2020,Selub:2021zdq,hogan2023causal,*hogan2024cosmic,Hogan_2023}. For example, the temperature anisotropy 2-point correlation function is conspicuously close to an exact zero at $90^\circ$ (within \:\!$\sim\:\!\!3\mu\mathrm{K}^2$), along with a strange dominance of odd multipoles over even ones in the angular power spectrum ($p<0.003$ at $\ell=28$)~\cite{Muir_2018,jones2023universe,Copi_2009,*Schwarz_2016,Hagimoto_2020}. The standard interpretation is to assume that these are just individual statistical anomalies, but interpreted in the context of possible symmetries on the horizon that might cause a conspiracy of interrelated anomalies, our universe looks exceedingly rare among \:\!$\sim\:\!\!10^6$ simulated realizations based on the standard QFT-based model, reaching $p$-values under $p<10^{-4}$ in several statistical tests~\cite{Hagimoto_2020,Selub:2021zdq,Hogan_2023,hogan2023causal,*hogan2024cosmic}. Given that this standard model was never founded on consistent, well-motivated principles\,--- with some clearly unphysical assumptions made, such as plane wave modes that are prepared acausally and ``freeze out'' acausally\,\cite{Hogan_2023,hogan2023causal,*hogan2024cosmic,Hogan2018a,Hogan_2020,Hogan2022_1}\,--- it would behoove the community to seriously consider alternatives. There may be signatures of quantum gravity that are easily reachable simply by updating models of inflation with coherent delocalized quantum states on causal horizons instead of local effective fields on classical backgrounds\,\cite{Hogan_2023,hogan2023causal,*hogan2024cosmic,Hogan2018a,Hogan_2020,Hogan2022_1}. An exciting new direction for possible research is the recent detection of parity violations in large-scale galaxy distributions at $2.9\sigma$ and $7.1\sigma$ by two independent analyses\,\cite{Hou_2023,Philcox2022}, which is difficult to account for using a QFT-based model of $\mathbb{P}$-symmetry violation\,\cite{Philcox2023} but could be a natural fit with the type of antisymmetry on causal horizons found in the 't~Hooft black hole model.

\enlargethispage{2.5pt}
In the following sections, we will construct candidate geometries, design parameters, and projected spectra for interferometric probes by putting together the most compelling elements from the literature reviewed above. The scaling of the holographic fluctuations, previously estimated from heuristic reasonings\,\cite{Kwon:2014,Hogan_2017,Hogan:2015b,Hogan:2016}, will adopt values consistent with the well-grounded results in the Banks-Zurek theory\,\cite{Banks_2021} and other Verlinde-Zurek models\,\cite{Verlinde_2021,Verlinde:2022hhs}. The former is based on a framework of conformal field theory and entanglement entropy, originally developed in the AdS/CFT context, but updated such that the scales of causal diamonds in flat space-time take on the role played by the AdS boundary scale. The latter are based on mapping fluctuations in black hole thermodynamics to flat space-time via a radially accelerated coordinate frame or summating the memory effect on gravitational shockwaves from vacuum fluctuations, and share the same geometric correlation functions as the 't~Hooft black hole models\,\cite{Hooft:2016itl,*Hooft:2016cpw,*Hooft2016_,Betzios_2016,tHooft2018,Hooft:2018gtw,*Hooft:2018syc,*hooft2019,Gaddam_2022}, but map the quantum operators onto definite scalar metric fluctuations using field theory. Our modeling of the geometric correlations, and more importantly the detector response, will refer to the original 't~Hooft algebra of the quantum states, to better project the effects of the quantum measurements being made of space-time. Lastly, the time correlations in the signal will be modeled based on the scaling of quantum information in the Banks-Fischler HST theory\,\cite{banks2023fluc,Banks2011,Banks:2018aed,Banks_2020,Banks:2020dus,*banks2022old,banks2023hilbert,*banks2023impossibility,*BanksGRF2023}.

\count\footins = 850
\section*{\NoCaseChange{Design Principles and Interferometer Geometry}}

The 't~Hooft and Verlinde-Zurek models provide much insight about what kind of quantum fluctuations\,--- and correlation modes thereof\,--- are to be expected on causal horizons. However, translating these theorized signatures to observables measured in a real experiment is a highly nontrivial problem. In a full theory of quantum gravity, merely identifying the operators corresponding to the observables would be sufficient, but neither of those models offer such a complete framework. The 't~Hooft algebra, Eq.\,(\ref{hooftcorr}), comes closest to identifying the actual quantum microstates that we would need for a rigorous calculation where those quantum geometric states can be coupled to the photon states and the measurement can be described by the appropriate quantum operators. However, it was derived on a Schwarzschild geometry, and it needs to be appropriately translated to a Hilbert space for flat space-time. Furthermore, despite its robust description of the quantum correlations that transversely extend across the entire horizon, the quantization is limited to the radial modes, and thus it struggles to implement the correct holographic statistical mechanics of the black hole, and does not give the correct magnitude of quantum fluctuations on the horizon. The Verlinde-Zurek model, Eq.\,(\ref{zurekcorr}), on the other hand, gets the right scale of horizon fluctuations for a causal diamond by utilizing the tools of quantum field theory (rather than 't~Hooft's choice to stick to classical quantum mechanics to obtain the full algebra), but in the process compromises the quantum nature of the correlations, reducing the phenomenology to classical correlations of definite scalar metric fluctuations. Despite the insightful Banks-Zurek derivation of holographic thermodynamics in flat space-time using the tools of entanglement entropy and conformal field theory, and the later Verlinde-Zurek reproduction using gravitational shockwaves, such QFT-based mappings are quite limited when it comes to calculating actual correlations. As the operator description of quantum geometric states is mapped onto definite perturbations on a background metric, the correlations rely on a Green's function that does not preserve the nonlocal, indeterminate character of the quantum correlations in the 't~Hooft $S$-matrix. While it is very compelling that these QFT-based mappings appear to reproduce the same spherical harmonic correlation functions for the space-time fluctuations\,--- and in some cases even originate from a similar set of quantum commutators\,--- as the 't~Hooft algebra, it should be stressed that when we try to model an actual measurement in quantum space-time by coupling the quantum geometric states to a concrete measurement, the detector response and the projected signal will not be the same. To more closely model the consequences of these measurement operations, our projections will reference the original 't~Hooft algebra for the physical characteristics of these quantum correlations. This will be especially relevant for the handling the transverse degrees of freedom, as the 't~Hooft and Verlinde-Zurek models are differently limited in handling these, and some inferences and conjectures will need to be made to reach a concrete a model projection.\hspace{3.5pt} 

We emphasize that local scalar metric fluctuations, taken literally, must correspond to degrees of freedom that scale extensively, which obviously cannot be holographic. If these correlation functions are correct, the fundamental physics of black hole-like causal horizon states must be described by the original operator algebra derived by 't~Hooft. The choice made in Verlinde-Zurek to map 't~Hooft's quantum operators onto metric fluctuations using field theory is a compromise made for better calculational control, but one that changes the physics quite significantly. It takes a quantum fluctuation that is nonlocal, indefinite, and observer dependent and turns it into a scalar metric fluctuation that is local, definite, and identical for every observer, which is a problem when we are trying to describe an infrared effect with holographic degrees of freedom. To avoid this issue, Zurek and colleagues have stated that the resulting mathematical formulae should be considered only to be valid for a specific degree of freedom that is defined for a specific measurement~\cite{LiPRD,Lee_2024,lee2024propertime}. As this manuscript primarily serves to present a new experimental research program, it aims to conjecture a broader picture that better captures the physical nature of these fluctuations even at the cost of some mathematical precision. For example, one might naively expect the Verlinde-Zurek metric correlations to be directly measurable in a Michelson interferometer that inscribes the causal diamond (Fig.\,\ref{diamond}), but the expression contains a logarithmic divergence as $|\mathbf{r}_1-\mathbf{r}_2|\rightarrow0$. This ``autocorrelation'' limit can be absorbed into the definition of $L$ for a macroscopic device where only the low $\ell$ modes correlated over larger transverse scales are detectable, but interpreted straightforwardly as metric fluctuations on causal diamonds, the steep difference in correlation power at fine angular scales would lead to a blurring of astrophysical images over cosmological distances, which is ruled out by existing data\,\cite{PerlmanRappaportChristiansenNg2015}, as will be elaborated upon later. This behavior is likely a consequence of the Verlinde-Zurek model only deriving a longitudinal effect and treating the transverse correlations with a cutoff, and as we will see, accounting for the transverse extent of the quantum correlations and taking 't~Hooft's coherent states on horizons seriously are key to obtaining a spectrum that matches primordial signatures in the CMB and preserves scale invariance at high $\ell$.\hspace{3.pt} 

\enlargethispage{4.5pt}
In our design, we will take seriously the 't~Hooft picture of coherent quantum states extended across the entire causal horizon. There is no definite effect until a specific measurement has been made, where the observable is subject to the specific scales and characteristics of the device (such as a telescope). The quantum indeterminacies are not mapped\linebreak

\begingroup
\setlength{\columnsep}{16.pt}
\begin{wrapfigure}{l}{.25\textwidth}
\vspace*{-11.75pt}
\includegraphics[width=\linewidth]{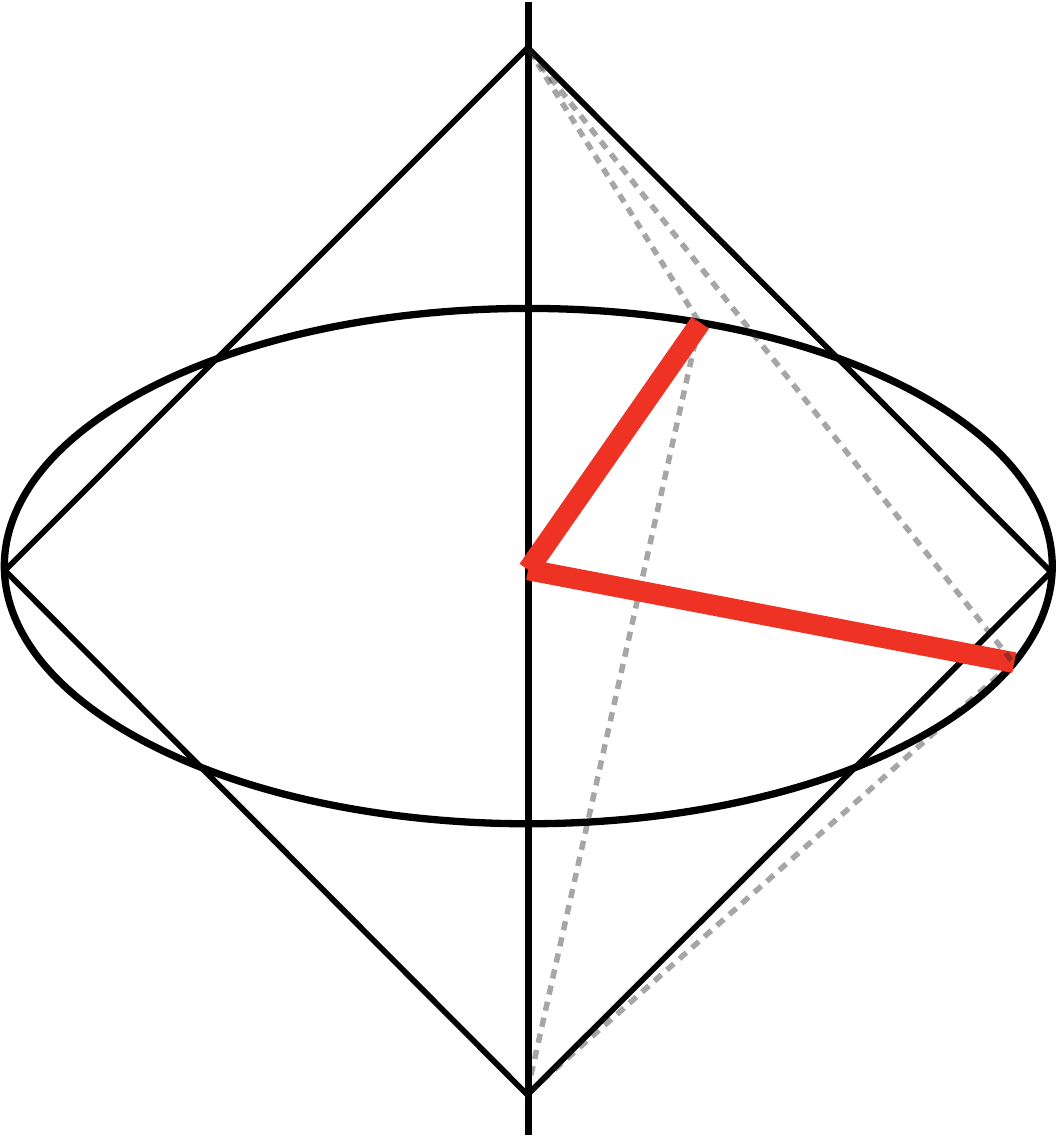}\vspace{-4.pt}
\protect\caption{Null trajectories inside a simple Michelson interferometer exactly trace the surface of a matching causal diamond.\label{diamond}}\vspace{-24.pt}
\end{wrapfigure}
\vspace{-12.5pt}\noindent onto local metric fluctuations. Because the eventual full theory must be independent of a classical definite background, we will insist that space-time cannot be measured without specifying an answer to the \!\!\;``relative to what?''\! question, just as a gravitational potential cannot be defined locally without a reference to another observer. This leads us to the first guiding principle of our experimental design:\vspace{6.125pt}
\begin{enumerate}[left=\parindent .. 2.167\parindent]
\item All causal horizons are universal boundaries of coherent quantum information, where the decoherence of space-time happens for the observer.
\end{enumerate}\vspace{6.125pt}
This is posited as an alternative to the Verlinde-Zurek postulate on the universality of metric fluctuations on horizons. Similar to Verlinde-Zurek, we insist that quantum uncertainties on the black hole horizon must analogously apply to causal diamonds in flat space-time (conformal Killing horizons), but retain 't~Hooft's coherent states on causal horizons instead of mapping them onto metric fluctuations. This picture is consistent with the increasing robust awareness that in solving the black hole information problem, the community needs to elucidate the nonlocality of quantum mechanical states on causal horizon scales, and that a quantum theory of space-time should start with an emergent formulation based on Hilbert spaces instead of local field theoretic perturbations on a background~\cite{Giddings_2004,*Giddings_2019,Giddings_2019_,*Giddings2020,*Giddings2022,Nomura:2012cx,Cao:2017hrv,Lowe_2006}. The background independence of space-time might be intricately related to its non-local realism.\hspace{2.5pt}

The idea of (de)coherence on general Killing horizons\,--- and not just black hole horizons\,--- fits in with several other compelling theoretical studies at the intersection of quantum mechanics and space-time. For example, the ``fast scrambling'' near causal diamond boundaries in the Banks-Fischler HST framework may be akin to transitions between coherent states and decohered ones near black hole horizons in the 't~Hooft model. Decoherence and scrambling are of course different processes, but their effects are interrelated, and the connection between them is an interesting topic for further research\,\cite{Lowe_2016,Touil_2021}. Another hint that an event horizon is not necessary for the quantum information decoherence usually associated with black holes comes from the fact that Hawking radiation itself does not actually require an event horizon and only needs an apparent horizon defined by a local observer~\cite{VISSER_2003,Wondrak_2023}.
\endgroup

In fact, the decoherence effect of horizons is not limited to the case of Hawking radiation, where one member of an entangled pair of particles crosses the horizon. Wald and colleagues have recently shown that a spatial superposition in the mere vicinity of a horizon undergoes decoherence because the horizon harvests ``which path'' information when gravitational radiation from the mass-energy of the quantum state enters the horizon~\cite{Danielson_2022,Danielson_2023}. This effect holds for both black hole event horizons and general Killing horizons (e.g. Rindler horizons), and in the latter case, is distinct from and larger than the decoherence from Unruh radiation. Of course, the decoherence here involves the quantum state of a particle and not of space-time itself, but the thought experiment should be interpreted as ``a quantum mechanical realization of [Wheeler's participatory universe], where space-time itself plays the role of the observer.'' This is again reminiscent of the quantum-classical boundary for information on the 't~Hooft black hole horizon. 

There are even further hints in studies of quantum foundations that this quantum-classical boundary for information on horizons might generally apply to Killing horizons formed by accelerated coordinate frames in flat space-time. For example, we have seen renewed interest in the Wigner's friend \textit{gedankenexperiment} in recent years, as analyses of a modified version involving Bell-type states showed that the quantum theory could not consistently describe the use of itself, resulting in a no-go theorem for observer-independent facts~\cite{Renner2018,Brukner2018}. However, a common loophole in all of the proofs is the assumption of a classical formulation of locality based on a definite background space-time. An intriguing manuscript by Durham, where the same problem was studied in the context of a Rindler horizon, again reveals the role of the horizon as a kind of quantum-classical boundary~\cite{Durham}. If the background fabric of space-time is stitched out of Hilbert spaces that are bounded by causal horizons that determine the localization and coherence of quantum geometric information, we might have a path towards resolving this paradox.

\enlargethispage{4pt}
As a corollary to the first principle, we propose a second postulate:\vspace{5.375pt}
\begin{enumerate}[resume,left=\parindent .. 2.167\parindent]\setcounter{enumi}{1}
\item The states of space-time are quantized on null surfaces\,--- black hole horizons, light cones, et cetera\,--- without a background. These surfaces set the symmetries of quantum geometry. The states have coherence scales matching the size of the causal boundary along its surface. However, along the world line\,--- e.g. between layers of parallel light cones or nested causal diamonds\,--- the states are independent and uncorrelated at the Planck scale~\cite{Banks_2021,banks2023fluc}. 
\end{enumerate}\vspace{5.375pt}

\begin{wrapfigure}{r}{.25\textwidth}
\vspace*{-2.pt}
\includegraphics[width=\linewidth]{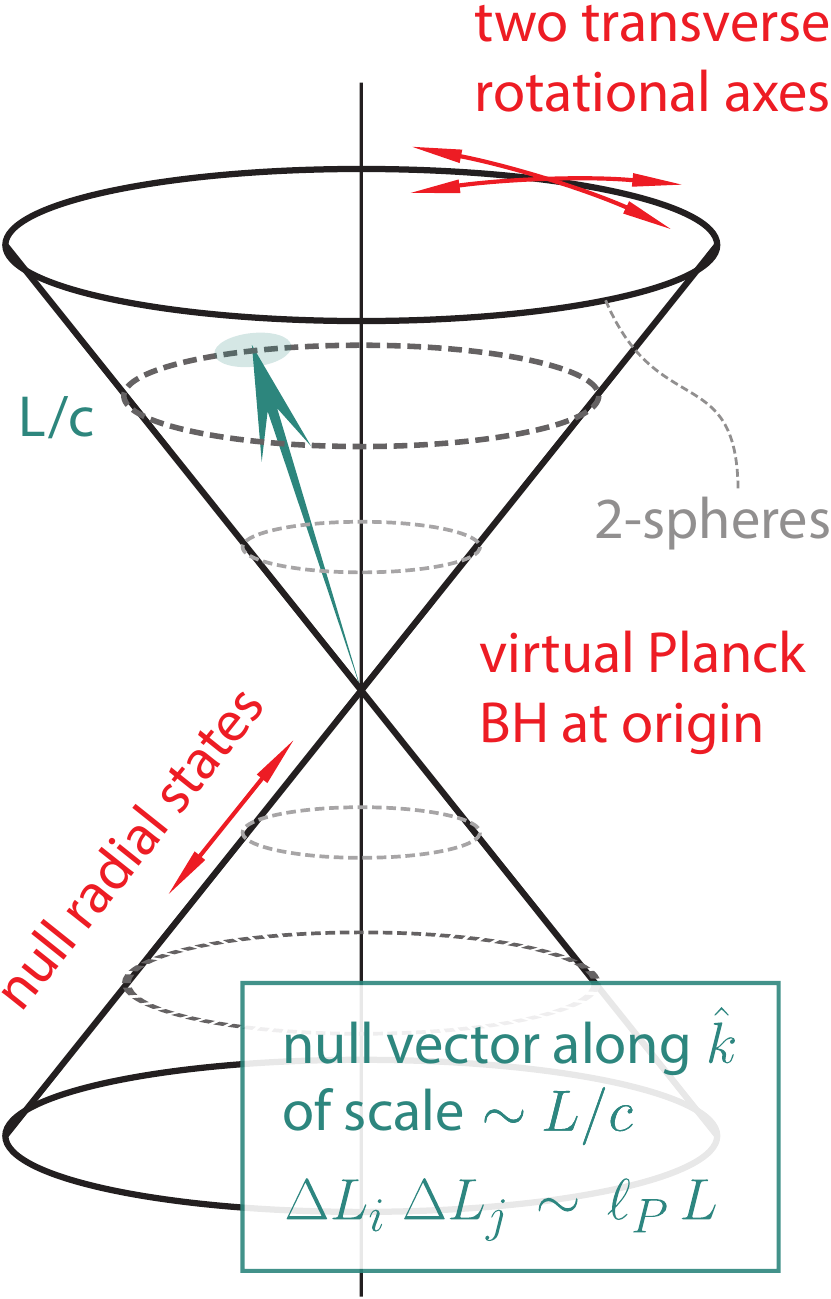}\vspace{-7.5pt}
\protect\caption{States of quantum geometry are to be quantized on the Dirac light cone, Eq.\,(\ref{diracdelta}), featuring the same 4D antisymmetry as in the 't~Hooft eternal Schwarzschild model, Eq.\,(\ref{hooftcorr}), which allows asymptotic mappings of null foliations if we have a virtual Planck black hole at the origin. The spherical symmetry of the system is posited to accept rotational solutions.\vspace{0pt}\label{lightcone}}
\end{wrapfigure}

This is our modern version of Wheeler's vision for a theory of space-time that captures the nonlocality of quantum mechanics by using light cones around the world lines of observers as the building blocks of background space-time itself\,\cite{WheelerQuote}.\footnote[3]{``Just as the proper recognition of this atomicity requires in the electromagnetic theory a modification in the use of the field concept equivalent to the introduction of the concept of action at a distance, so it would appear that in the gravitational theory we should be able in principle to dispense with the concepts of space and time and take as the basis of our description of nature the elementary concepts of world line and light cones.''\vspace{2.25pt}} It also reflects certain lessons we learned from the first-gen Holometer. That instrument was designed on the naive premise that we might be able to measure some noncommutativity of space-time in a simple Michelson setup similar to Fig.\,\ref{diamond}, based on a toy model of holographic dimensional reduction. Since the detector measured translational degrees of freedom along two spatial dimensions, the proposal faced much (justified) criticism that the targeted effect was not Lorentz invariant from colleagues who recognized similarities to other models of noncommutative geometry (which are usually implemented perturbatively on a classical field theoretic background). As the trajectory of light inside the interferometer traces the boundary of a causal diamond, the null result from this early experiment should be thought of as a verification of an exact symmetry, and tells us that we should probe effects that preserve this structure going forward. In a conjectured framework in which space-time is built out of certain quantum ``building blocks'' without a background, a model is manifestly Lorentz invariant by construction if it is built out entirely of Lorentz invariant objects, and use it to derive a Lorentz scalar.\footnote[6]{As previously mentioned, the usage of effective field theory in models by Zurek and colleagues has attracted some misconceptions, because those models are obviously not background independent, and the calculation involves specific reference frames. These approaches should be seen as attempts to rigorously circumvent the foundational issues in QFT while still retaining their utilization as powerful calculational tools. Taken too literally, the results can often be misinterpreted as violating Lorentz invariance, but Zurek has taken specific care to clarify that her results should only be taken literally for a specific observational degree of freedom carefully defined in the context of a specific measurement in a way that maintains Lorentz invariance\,\cite{LiPRD}. A more in-depth discussion of the subtle issues involving the gauge degrees of freedom of quantum space-time is outside the scope of this manuscript, but there has also been recent work demonstrating the gauge invariance of the interferometric observable in these types of experimental probes\,\cite{lee2024propertime}, as well as a derivation of covariant Noether charges in a spherically symmetric causal diamond in flat space-time that may correspond to this phenomenology\,\cite{bub2024Noether}.} While we are positing a quantum effect that is dependent on a specific measurement by an observer, there is no point at which a frame of reference enters the calculation. Our current generation of models are designed such that the correlations are built out of quantum objects that ``live on'' causal surfaces, and the measured value of the experimental observable is obviously a scalar. The conjectured quantum fluctuations preserve the shape of the light cones, just like Lorentz boosts and rotations do (Fig.\,\ref{lightcone}).

If each causal diamond layer has Planckian ``thickness'' along the world line, and has an independent Planck scale fluctuation that is coherent along its surface, how do we design a geometric configuration that maximizes the accumulation of these fluctuations? In Banks-Zurek\,\cite{Banks_2021}, the authors adopt a model of nested causal diamonds, starting from the origin and growing along the world line by a Planck time for each successive layer until it reaches the system size of the interferometer. But we do not actually measure all of those different-sized layers. A single-shot measurement of the largest diamond delimiting the measurement would presumably contain all of the accumulated uncertainty, but what is the diamond fluctuating relative to? In a theory without a background, a single-shot measurement simply defines what the space-time is. In any case, we cannot make such a single-shot measurement of sufficient sensitivity.\hspace{2.5pt}

Of course, an interferometer also only measures differential fluctuations over time, not absolute lengths one can compare against an expected value. Any given measurement just establishes a nominal value of measured phase, and each successive measurement gives a relative phase compared to the earlier measurement. Each photon entering the interferometer traces the surface of an equally sized causal diamond delimiting the measurement, at different times. If the time offset $\tau$ is smaller than the total duration $2L/c$, the causal diamonds partially overlap. The overlapping volume of space-time must be consistent between the two measurements, and the non-overlapping interval can potentially contribute to differential fluctuations if coupled to the right observable (boundaries offset by $N$ Planckian layers have differential uncertainty that scales as $\sqrt{N}$\,). If $\tau$ is larger than $2L/c$, the two measurements are causally disconnected and independent. This means that for maximum quantum indeterminacy, or noise power, we want each measurement to pass through as many Planckian layers of causal surfaces as possible, and for maximum statistical power, we want the total integration time to contain as many independent samples of duration $2L/c$ as possible.

We have so far discussed the quantum geometric information available, but what part does an actual detector access? How does the real observable\,--- measured photons\,--- couple to quantum space-time in constituting a measurement?\hspace{3.5pt}

As stated above, a straightforward interpretation of Verlinde-Zurek\,\cite{Verlinde_2021} as local metric fluctuations runs into significant phenomenological issues. However, while Verlinde-Zurek demonstrated mathematically that the correlations on the 't~Hooft black hole can also be modeled in flat space-time, they diverged conceptually from the way 't~Hooft intended his model to be understood\,\cite{Hooft:2016itl,*Hooft:2016cpw,*Hooft2016_,Betzios_2016,tHooft2018,Hooft:2018gtw,*Hooft:2018syc,*hooft2019,Gaddam_2022}. Here, we will try to recover some of 't~Hooft's original interpretation of the quantum states, and combine it with some of the lessons from our phenomenological work to infer the observables and measurements which may give us irreducible quantum uncertainties.

\setlength{\columnsep}{16.pt}

To do so, we will borrow from Dirac's old approach to writing down quantum commutation relations in a Lorentz covariant way. Dirac sets up the following modified $\delta$-function to define localization on light cones~\cite{DiracQM}:\vspace{-2.5pt}
\begin{equation}
\mathbbl{\Delta}(x) = 2\delta(x_\mu x^\mu)x_0/|x_0|\qquad\qquad\hspace{0.5em}\mathbbl{\Delta}(-x)=-\mathbbl{\Delta}(x)\label{diracdelta}\vspace{-2.5pt}
\end{equation}
This is a $\delta$-function that vanishes everywhere except on the light cone, but unlike the typical $\delta(x_\mu x^\mu)$, has a sign flip between future and past light cones. It is even at spacelike separations but odd in timelike directions, which results in a 4D point-parity antisymmetry that combines time and space. Dirac uses this function to establish covariant commutation relations for field operators: $[\hat{A}_\mu(x),\,\hat{A}_\nu(x')]=g_{\mu\nu}\,\mathbbl{\Delta}(x-x')$.

A common way of understanding space-time fluctuations at the Planck scale is to think of the virtual states popping in and out of the vacuum as small Planck sized black holes. Fig.\,\ref{lightcone} is a schematic representation of a Dirac light cone, where we may imagine at the origin a Planck sized virtual black hole as modeled by 't~Hooft. If we insist that quantum space-time has to be quantized on this covariant boundary, the states constrained to this structure follow the same 4D antisymmetry as the 't~Hooft black hole states in Eq.\,(\ref{hooftcorr}). This means that if we match the null cone foliation in the rest frame of the Planck black hole to the null cone foliation in distant asymptotically flat space-time, we could establish a form of equivalence principle, where the holographic correlations have the same effect on states of light tangent to a light cone in flat space-time as they do on tangential components of field states near a black hole horizon. Of course, one also expects this from considerations of generalized horizon entropy and Verlinde-Zurek's success using topological black hole transformations, but the difference here is that we seek to preserve 't~Hooft's nonlocal coherent quantum states with antipodal antisymmetry. As 't~Hooft asserts\,\cite{tHooft2018}, ``If it is true that vacuum fluctuations include virtual black holes, then the structure of space-time is radically different from what is usually thought.'' But perhaps our modern paradigms just needed to return to the ideas that Wheeler and Dirac prefigured\,\cite{WheelerQuote,DiracQM}.

We now state two hypotheses about the specific quantum states and the observables that may lead to a detectable effect. One is straightforwardly adopted from the works cited above; one will rely on some intuitive reasoning:\vspace{4.125pt}
\begin{enumerate}[resume,left=\parindent .. 2.167\parindent]
\item The coherent quantum geometric states on causal diamonds have a spherical harmonic expansion. Due to frame dragging, the entire null horizon is a single quantum object with nonlocally extended states. Its relation to incoming\,/\,outgoing information resembles a hydrogen atom, as described by the 't~Hooft gravitational $S$-matrix.\vspace{3.375pt}
\end{enumerate}

\begin{enumerate}[resume,left=\parindent .. 2.167\parindent]
\item As expected for a system with spherical symmetry spanned by spherical harmonic states, the quantum space-time system accepts rotational solutions. An irreducible quantum uncertainty from holographic dimensional reduction is measured via incompatible observables\,--- e.g. transverse fluctuations around two orthogonal rotational axes.
\end{enumerate}\vspace{4.125pt}

\enlargethispage{2.5pt}
Why do we propose measuring rotational fluctuations? Let us first consider the radial fluctuations (along surfaces of proper time). As noted before, a straightforward interpretation of Eq.\,(\ref{zurekcorr}) as local metric fluctuations would lead to a blurring of astrophysical images over large distances, which is ruled out by data\,\cite{PerlmanRappaportChristiansenNg2015}. Local metric fluctuations are likely not the correct interpretation\,--- in a model where space-time and locality are emergent from relationships between macroscopic quantum states and observers, predictions of scalar gravitational potential fluctuations are quite risky, since they are the same in every frame. The angular correlations are likely from coherent nonlocal quantum states in the 't~Hooft interpretation, rather than being described by a Green's function in a local framework as in Verlinde-Zurek. In this case, subtleties about the specific measurements must be considered. The naive approach would be to make longitudinal measurements of null radial fluctuations, say, using the directional arms of a Michelson interferometer, with some variation in the geometry to modulate angular correlations\,--- e.g. an interferometer where the angle between the two arms can be changed, or two cross-correlated interferometers where one is offset by a varying angle relative to the other. In this approach, the first-gen Holometer, with a simple Michelson geometry, already constrains the Verlinde-Zurek spectrum in Eq.\,(\ref{zurekcorr}) to a normalization prefactor of $\alpha\lesssim0.7$ at the $3\:\!\sigma$ level~\cite{LiPRD}. This first-gen Holometer experiment was originally designed on the premise of non-commuting directions. But in the 't~Hooft spherical harmonic states, it is unclear that there is anything non-commuting between different directions. Radial measurements in two different directions might simply result in the system ``collapsing'' to the same state, since the two radii can be correlated on the same spherical harmonic state (unlike the case of local metric fluctuations). Typically, making a continuous measurement of the same quantum system with the same observable results in the Zeno effect, where the system never relaxes away from the measured state.\footnote[5]{While this paper is written implicitly in the language of a modern understanding of the Copenhagen interpretation, in other interpretations (e.g. Everett), there is no discrete state update or projection, and the measurement is described as just a correlation (applying the Born rule with the measuring device included in the analysis)\,\cite{Wallace2019}. It is not obvious how the Zeno effect should be handled in the context of the measurement problem in quantum space-time, given the challenges of identifying the relevant states of space-time to use in the analysis, at least until a signal is detected. However, regardless of the foundational perspective, the Zeno effect is a real phenomenon, verified by empirical tests. The relevant factor here in avoiding the Zeno effect is whether the timescale on which the measurement is repeatedly carried out is sufficiently long relative to the timescale on which the measured system evolves. If one posits that the latter is Planck time for a system of quantum space-time, there is no Zeno effect, since the former is set by the detector response, much slower than Planck time. But such a hypothesis would lead to a stochastically evolving quantum space-time background that has no relation to any measurement thereof whatsoever, which is problematic in the context of satisfying holography in an arbitrary causal diamond in flat space-time and avoiding the CKN infrared catastrophes. In the frameworks adopted here, the coherence scale (or relaxation/decay time) of the quantum space-time is the timescale of the causal diamond (or 4-volume of space-time) being measured, in which case the timescales of the interferometric measurements are much shorter (which is the point of designing the detectors with fast response times).}

In an interferometer, each photon is split by a beamsplitter and exists in a delocalized superposition across two arms until the interference is measured, in which case we can reasonably assume that the geometries of both arms would couple to a single photon state. Ideally, we want to design an interferometer where each measurement couples to two incompatible space-time observables. However, for the radial fluctuations, this may not be possible; it may be that two observers are required. After all, the gravitational potential cannot be defined locally without a reference elsewhere. Furthermore, unlike in the toy model of Ref.\,\cite{krishogan}, in empty space-time, there is no source generating the fluctuations; there are only virtual states that pop in and out of existence. Such effects have zero mean and nonzero mean square power\,--- the energy and information are being exchanged relationally. All observers and physical systems create their own space-time without a background, comprising these relations, woven together in a way that respects consistency conditions for overlapping causal volumes, perhaps as in the Banks-Fischler HST framework. This picture lends itself naturally to the 't~Hooft gravitational $S$-matrix approach. One observer's causal diamond consists of a outgoing light cone of information as well as an incoming light cone of external information from other observers, from the larger Hilbert space outside one's own Hilbert space. How do we detect irreducible quantum uncertainties in this model? In the 't~Hooft model, the nonzero commutator is between incoming and outgoing modes at opposite directions (Fig.\,\ref{lightcone}). It is very difficult to design a single interferometer geometry that measures the two concurrently. If we may relax the requirement of what constitutes a ``concurrent'' measurement of two incompatible observables, we could consider using two interferometers to test the 't~Hooft commutator, with the quantum uncertainty manifesting \textit{between} the two measurements (as opposed to being shared in their cross-correlations). However, unlike the case of metric fluctuations, it is unclear that such a measurement by two different observers with their own causal diamonds would show irreducible uncertainties in quantum space-time. A detailed such study investigating couplings between quantum space-time and the states of different detectors and their photons is left to future work.

An alternative candidate pair of non-commuting operators is between incoming and outgoing null modes that together close a causal diamond (Fig.\,\ref{diamond}), as in the gravitational shockwave interpretation of the Verlinde-Zurek model (before the fluctuations are mapped onto metric fluctuations). In this case, a photon making a round trip in any interferometer arm would make a measurement that combines both modes. But the uncertainty would be in the size of a \textit{single} causal diamond, defined by the point of reflection at the end mirror of the interferometer arm. An individual measurement would simply define the nominal value for the size of that diamond in a background independent system of quantum space-time, as explained in our earlier discussion of a single-shot measurement. To measure an accumulating uncertainty, we must make successive differential measurements relative to this reference value. Such measurements are perhaps made, over many causal diamonds corresponding to photons injected into the interferometer over time. However, as each of these photons couples to the sum of the incoming and outgoing null positions (not   directly measuring either individually), each measurement may actually result in a partial ``collapse'' of the quantum states afterwards, in the degree of freedom corresponding to this sum\,--- and the measurement does not fully access the non-commutativity in this theory, which is enforced at the end mirror reflection. Hence, the successive differential measurements may not be immune to the Zeno effect discussed above. This may sound confusing in the context of a permanent gravitational memory that accumulates the Planckian shockwaves, but the point is that these individual shockwaves are quantum states until measured. As established earlier, to maximize the measured uncertainty, we want \textit{each} measurement to contain and couple to \textit{different} causal diamonds that are macroscopically separated by as many Planckian layers of causal diamonds as possible in between (to capture the differential fluctuations, ``collapsing'' all the layers in between), which is easy when part of the photon path in the interferometer arm is bent (shown later in Fig.\,\ref{IFOcorr}) but not possible with a photon making a simple radial round trip in a straight interferometer arm. For our model to avoid this issue and predict accumulating quantum uncertainties in straight-arm Michelson interferometers, we may need to make an additional assumption of continuous stochastic generation of new quantum geometric noise, e.g. by modeling the quantum state on each Planckian null foliation layer as having a Planck scale relaxation time.\hspace{6.5pt} 

Measuring rotational fluctuations could offer a way to avoid to these tricky issues. We expect a spherically symmetric system expanded into spherical harmonic modes to naturally accept rotational solutions. In the context of a space-time that is emergent without a background, we expect there to be an indeterminacy of direction, as the observer's inertial reference frame must also be emergent (the angular momentum uncertainty of any Planckian quantum state is inherited by the space-time through frame dragging\,\cite{Hogan:2015b}). The 't~Hooft ``hydrogen atom'' model of a black hole\,--- which can be mapped onto causal diamonds in flat space-time\,--- gives us much motivation to explore this possibility. Consider a picture where we have the orbital model of an atom, but instead of the wavefunction representing local probability densities of electrons, it represents (heuristically) quantum fluctuations in the local gravitational potential or metric. A local curvature fluctuation (again, loosely speaking) at a particular point on a causal horizon layer should cause phase shift effects in all directions, including both radial and transverse directions (of course adjusted for the mapping from a black hole horizon to causal diamond layers in flat space-time). This means that for each spherical harmonic mode, there should be angularly correlated transverse shifts as well as radial shifts. Just like the spherical harmonics of an atom correspond to rotational eigenstates of the electron cloud, the spherical harmonics of a causal horizon should contain similar rotational solutions. In the 't~Hooft black hole model, because only the radial components were quantized, angular filters are used at scales where the transverse gravitational effects become significant so that the system respects holographic total entropy. Quantum geometric uncertainties in (incompatible) transverse directions, angularly correlated by coherent states extended across the horizon, provide a natural mechanism for how this Planckian transverse mode cutoff (and dimensional reduction) is physically implemented in the system.\hspace{7pt} 

Notably, the transverse fluctuations do not have the logarithmic divergence issue at small angles that invalidate the radial ones, for two separate reasons. Wavefronts longitudinally distorted or corrugated by $\langle\hspace{.07em} \Delta R^2\hspace{.07em}\rangle\sim\sqrt{\ell_PR\hspace{.07em}}\sim10\,\mu\mathrm{m}$ from sources $R\sim1\,\mathrm{Gpc}$ away will not form clear astrophysical images in a telescope, because \textit{longitudinal} differentials between an aperture area versus a subarea (say, 1/10 of the radius) \textit{transversely} and directionally blur the focusing of the image.\footnote[1]{Zurek and colleagues have issued a rebuttal since an early preprint of this manuscript~\cite{Lee_2024}, but without addressing the defocusing effect. Their work controls the logarithm at small scales with a cutoff at the aperture scale, which the present author finds not well-justified.\hspace*{10.5pt}} An actual transverse fluctuation of $\langle\hspace{.07em} \Delta R_\perp^2\hspace{.07em}\rangle\sim10\,\mu\mathrm{m}$, on the other hand, is undetectable: far smaller than the transverse width of the wavefront from the optical diffraction of X-rays or $\gamma$-rays, or the aperture that is focusing the light. In addition, the signal from the transverse segments of the photon path (say, in an interferometer with a portion of the arms bent) represents a differential measurement between causal diamonds separated roughly by the scale of the transverse extent of the system (as we will see later in Fig.\,\ref{IFOcorr}), so the accumulated uncertainty scales with this transverse separation. In our model of an interferometric apparatus coupling to holographic noise, the logarithm at small angles in Eq.\,(\ref{zurekcorr}) is still present, but the measured fluctuation scales with the small time interval over which the photon trajectory is coupled to this logarithmically large correlation. The divergence at small angles is regulated by this added linear factor of the transverse extent of the correlation, and we are able to obtain a well-controlled calculation for the projected spectrum without introducing a cutoff as in the Verlinde-Zurek theory.

\enlargethispage{11.5pt}
It is possible that a viable model of radial fluctuations still exists. For example, if the modified angular power spectrum in our model for transverse fluctuations also applies to radial fluctuations, and if systems of quantum space-time are assumed to be constantly generating new noise in a stochastic manner (say, due to a Planck scale relaxation time between null foliations of space-time measurements), it may be that we can simply use straight-arm Michelson interferometers with varying angles, without the issues we laid out above for differential measurements of a background independent quantum space-time. As will be derived in the next section, in our model, the added scaling factor of the transverse extent of the coupling means that the angular spectrum of \textit{measurable} fluctuation powers distributed across the spherical harmonic modes scales as \:\!$\sim\:\!\!1/\ell^2$, instead of the \:\!$\sim\:\!\!1/\ell$ scaling we get if we simply sum over the $m$ modes in Eqs.\,(\ref{hooftcorr})~and~(\ref{zurekcorr}), even though we start with the same spherical harmonic eigenstate expansion. Of course, this is not surprising, given that Eqs.\,(\ref{hooftcorr})~and~(\ref{zurekcorr}) both only calculated radial modes. Even in the gravitational shockwave version of the Verlinde-Zurek model, the authors adopt the approximation of near-horizon planar light sheets and apply a cutoff in the transverse direction. It is quite probable that when the transverse directions are properly handled, including the quantum gravitational effects, there are no sharp peaks in correlation power at small angular scales, and no conflict exists with astrophysical imaging data. For example, the toy model in Refs.\,\cite{krishogan,mackewicz2024}, which uses classical general relativity to calculate the gravitational shockwaves of particles being emitted in indeterminate directions from isotropic decay events, shows the same \:\!$\sim\:\!\!1/\ell^2$ scaling derived here. As we have detailed in follow-up work\,\cite{hogan2023angular}, this \:\!$\sim\:\!\!1/\ell^2$ angular power spectrum is well-motivated from simple arguments of scale invariance, and is also consistent with a heuristic model of holographic primordial fluctuations on the inflationary horizon designed to match the angular correlations found in the CMB\,\cite{Hogan_2023}, if the correlations are mapped to flat space-time by removing the conformal expansion (going from comoving coordinates to normal coordinates). These four separate arguments give strong credence to the \:\!$\sim\:\!\!1/\ell^2$ spectrum presented by our model in the next section. The feasibility of an experiment probing radial fluctuations with this angular scaling will be considered in future work.

In this work, we will aim to identify designs where rotational degrees of freedom are expected to show irreducible quantum geometric uncertainties.  A natural way to do so is to draw an analogy with the angular momentum algebra for the atomic orbital, and couple the detector to two incompatible observables associated with non-commuting operators: transverse fluctuations with respect to two orthogonal rotational axes in three dimensions. In hindsight, this explains why the second-gen Holometer (a Michelson setup in a bent-arm planar layout) did not see a signal: in going from two orthogonal translational observables (first-gen Holometer, a straight-arm Michelson) to a rotation in two dimensions, we ended up with only one physical degree of freedom being measured. If the design proposed here detects an effect, we expect the magnitude to be of similar order as the radial effect suggested by Verlinde-Zurek, since both should be derived from the same physical phenomena and mechanism in flat space-time. Perhaps there are nuances of this physical system that can affect the specific shape of the projected time-domain correlation, but as we will see later, the frequency spectrum of the expected signal is fairly insensitive to subtle differences at this level. Since we are seeking the equivalent of the Bohr hydrogen atom model for quantum gravity, we should be able to reach projections that are serviceable for experimental design.

We present here candidate Michelson interferometer geometries (Fig.\,\ref{configs}) with both arms bent, where each arm measures one rotational degree of freedom respectively and the two rotational axes are orthogonal to each other. As will become clear in the spectra projections, the transverse propagation in the bent segment allows each photon to make a differential measurement between macroscopically separated causal diamonds (Fig.\,\ref{IFOcorr}). Importantly, photons inside an interferometer, once they enter through the beamsplitter, are delocalized across both arms as superposition states until measured at the output port (at near-perfect destructive interference). Each photon is coupled to the geometries of both arms ``at the same time,'' forcing a true concurrent measurement of two incompatible observables. To optimize the detector coupling to correlated fluctuations, we want the space-time being measured in the two arms to be separated by an angle (with respect to the beamsplitter) that maximizes the spherical harmonic correlations. For example, if we imagine (heuristically) two tracer photons traveling concurrently through the two arms of the configuration in Fig.\,\ref{configs}\;\!(a), the angle between the tracer photons is in the 45\,--\,60$^\circ$ sweet spot of Fig.\,\ref{angcorr}, and for Fig.\,\ref{configs}\;\!(b), the analogous angle is in the 0\,--\,30$^\circ$ sweet spot. In calculating this angular correlation function in Fig.\,\ref{angcorr}, we have taken only the odd harmonic modes of the correlation function $\mathbf{G}(\mathbf{r}_1,\, \mathbf{r}_2)$ in Eq.\,(\ref{zurekcorr}), without the $\ell=1$ mode. In general, we expect the $\ell=0$ and 1 modes to be unobservable in a local measurement: $\ell=0$ corresponds to the average scalar curvature of the system, and $\ell=1$ simply defines the inertial reference frame of the emergent space-time background ($\ell=0$ is the only mode with a nonzero mean, and $\ell=1$ is the only mode with a nonzero directional average, in the sense of being degenerate with a global displacement of the observer in a background independent system).

\section*{\NoCaseChange{Numerical Model and Projected Spectra}} 

\begin{figure}[t!]
\vspace*{4pt}\centering
\includegraphics[height=.3\textheight]{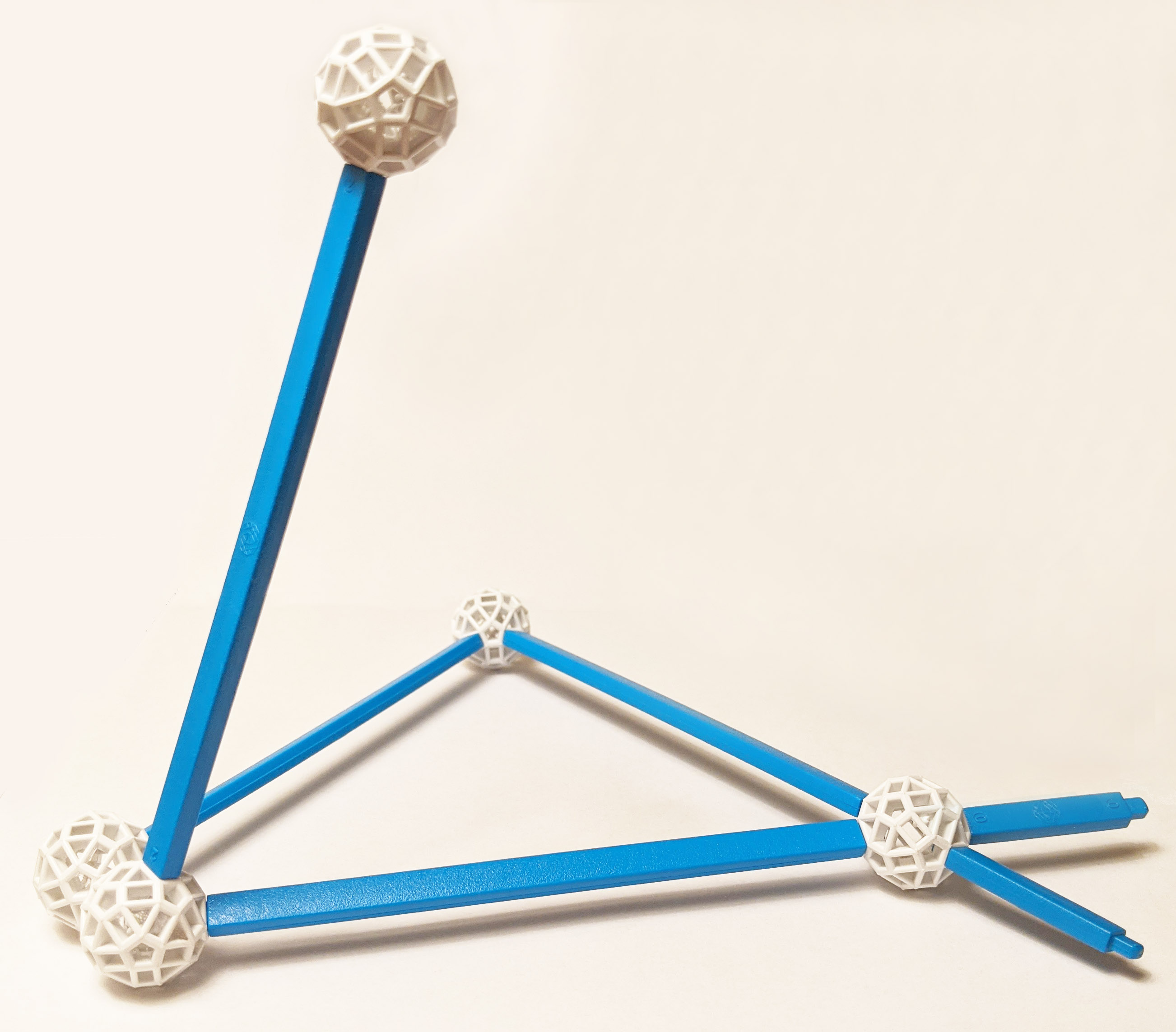}
\qquad\quad
\includegraphics[height=.3\textheight]{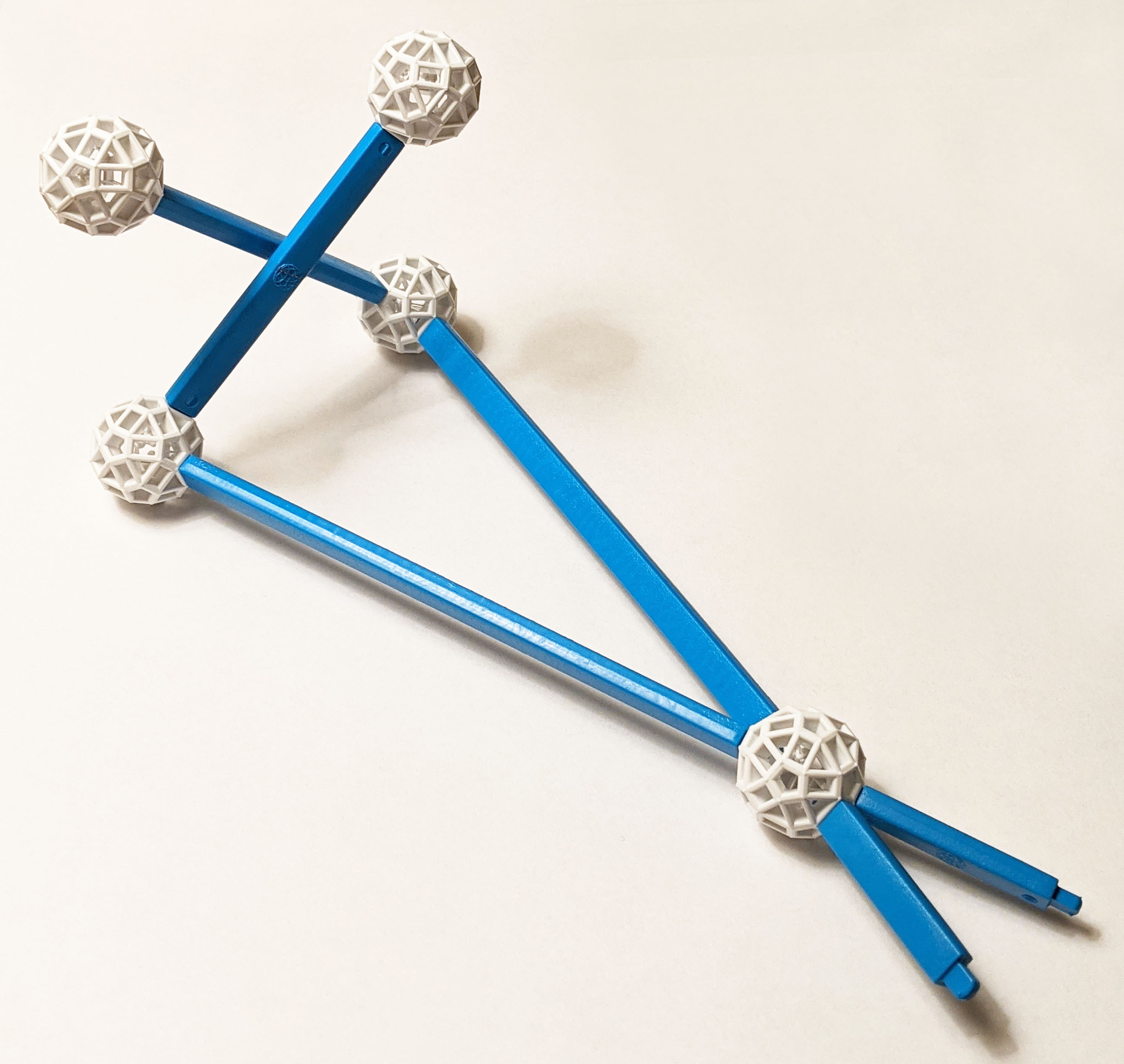}\hspace*{4pt} \\
\vspace{-195pt}\hfill(a)\qquad\qquad\qquad\qquad\qquad\qquad\qquad\qquad\qquad\qquad\qquad\quad\hspace*{3pt}(b)\qquad\hspace*{8pt}\vspace{183.5pt}
\caption{Interferometer geometries designed to optimize detector coupling to irreducible quantum geometric noise. (a)~All angles are $60^\circ$ and all segments equal. The horizontal plane of one arm and the vertical plane of the other intersect at $90^\circ$. (b)~The two bent segments intersect at $90^\circ$ at their midpoints, and the center of the bent segments is exactly orthogonal to the radial vector from the beamsplitter. The ratio between the bent segments and the radial segments is 2:3 (not depicted to scale).}
\label{configs}
\end{figure}

The design principles can be used to construct a numerical model for estimated signals. We will use configuration~(a) in Fig.\,\ref{configs} as an example, and the same calculation will apply to configuration~(b), which will be used as the nominal 3D configuration in a future experiment at Cardiff University\,\cite{grote2020,QIPublic}. 

\begin{figure}[t!]
\centering
\includegraphics[height=.25\textheight]{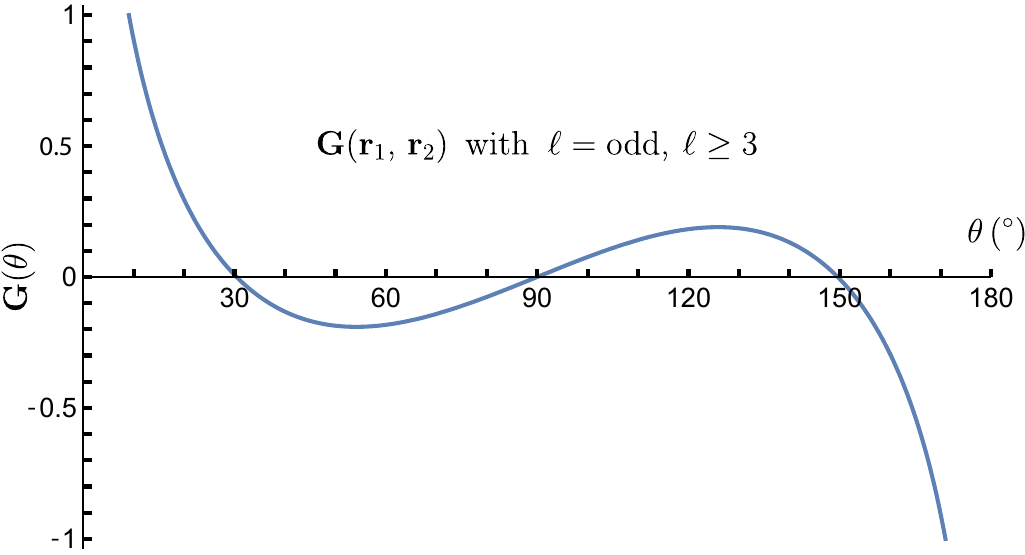}\vspace{-3pt}
\caption{The angular correlation function $\mathbf{G}(\mathbf{r}_1,\, \mathbf{r}_2)$ in Eq.\,(\ref{zurekcorr}), plotted versus the angle $\theta$ between $\mathbf{r}_1$ and $\mathbf{r}_2$, with the correction that we only count the odd spherical harmonics, starting with $\ell\ge3$. We identify two sweet spots to explore in this sum of Legendre polynomials: the positive correlations in the 0\,--\,30$^\circ$ range and the negative correlations in the 45\,--\,60$^\circ$ range. Note that the CMB temperature anisotropy 2-point correlation features the same antisymmetry and exact zero correlation at 90$^\circ$\,\cite{Hagimoto_2020}, even though it features significantly different physics due to the inflationary mechanism for its emergence.\,\cite{Hogan_2023,Hogan2018a,Hogan_2020,Hogan2022_1}. This correlation function applies to quantum uncertainties in space-time itself and should not be confused with the correlation predicted for a Michelson interferometer with an angle $\theta$ between the arms (e.g. Fig.\,2 of Ref.\,\cite{LiPRD}). For example, if the two arms of a standard Michelson interferometer had quantum geometric noise that is completely uncorrelated at 90$^\circ$, the differential arm length measured at the output would actually show the two noises added in quadrature.\vspace{-2.5pt}}
\label{angcorr}
\end{figure}

\enlargethispage{8pt}
An interferometer is sensitive to the phase difference between light that traveled through its two arms. Due to the conventions of gravitational wave interferometry (where the light travel time is much shorter than the sampling time of the output readout), the measurements are typically characterized by the differential arm length, $\mathrm{DARM}\,(\mathscr{t})\equiv [S_2(\mathscr{t})-S_1(\mathscr{t})]/2$, where $S_{2,1}(\mathscr{t})$ are the optical path lengths of the two arms measured by a photon that reached the detector at time $\mathscr{t}$. The time-domain correlation function for signals measured an offset interval $\tau$ apart is:\vspace{-.5pt}
\begin{equation}
\mathrm{DARM\;Corr}\,(\tau)\,\equiv\, \big\langle \mathrm{DARM}\,(\mathscr{t})\ \mathrm{DARM}\,(\mathscr{t}+\tau) \big\rangle \,\equiv\: \frac{1}{4}\,\sum_{\alpha,\,\beta}^{1,\,2}\,(-1)^{\alpha+\beta}\;\big\langle S_\alpha(\mathscr{t})\;S_\beta(\mathscr{t}+\tau) \big\rangle\vspace{-.5pt}\label{DARMcorr}
\end{equation}
In a realistic experiment\,\cite{grote2020}, we will likely use the cross-correlation between two identical interferometers that are nearly colocated and coaligned. This provides two independent measurements of quantum fluctuations in the same macroscopic space-time.\enlargethispage{3pt} If the instruments are carefully isolated such that there is minimal correlated systematic uncertainty, the remaining noise\,--- e.g. incoherent photon Poisson noise\,--- can be averaged down as $1/\sqrt{n}$ over large sets of data, while retaining the shared quantum space-time signal in the cross-correlation. It is important to note that both interferometers must each measure the same \textit{pair} of incompatible observables in order to detect an irreducible quantum geometric uncertainty that is correlated between the two detectors. In a configuration where two differently shaped interferometers were each measuring one out of a pair of incompatible observables, even if there was an irreducible uncertainty, the two observables would not be correlated. This would be akin to having two orthogonally oriented detectors measuring the spin of an electron at the same time; the measurements induce irreducible quantum uncertainty, but obviously the two measurements are not correlated at all. In such a configuration, the independent quantum geometric fluctuations would be below the noise floor of an individual shot-noise-limited interferometer, and a cross-correlation measurement would average down both the quantum space-time fluctuations and the photon quantum noise. It may be possible in the future, as some experiments are exploring\,\cite{McCuller:2022hum,vermeulen2024}, to achieve sufficiently low noise floors in a single interferometer by utilizing new technologies, such as single photon detectors (to eliminate shot noise) and cryogenics (to reduce thermal noise), but even with such advances it is likely that dual interferometers are necessary for control over systematics and confidence in a detection. 

\begin{figure}[t!]
\centering
\includegraphics[width=.4954\textwidth]{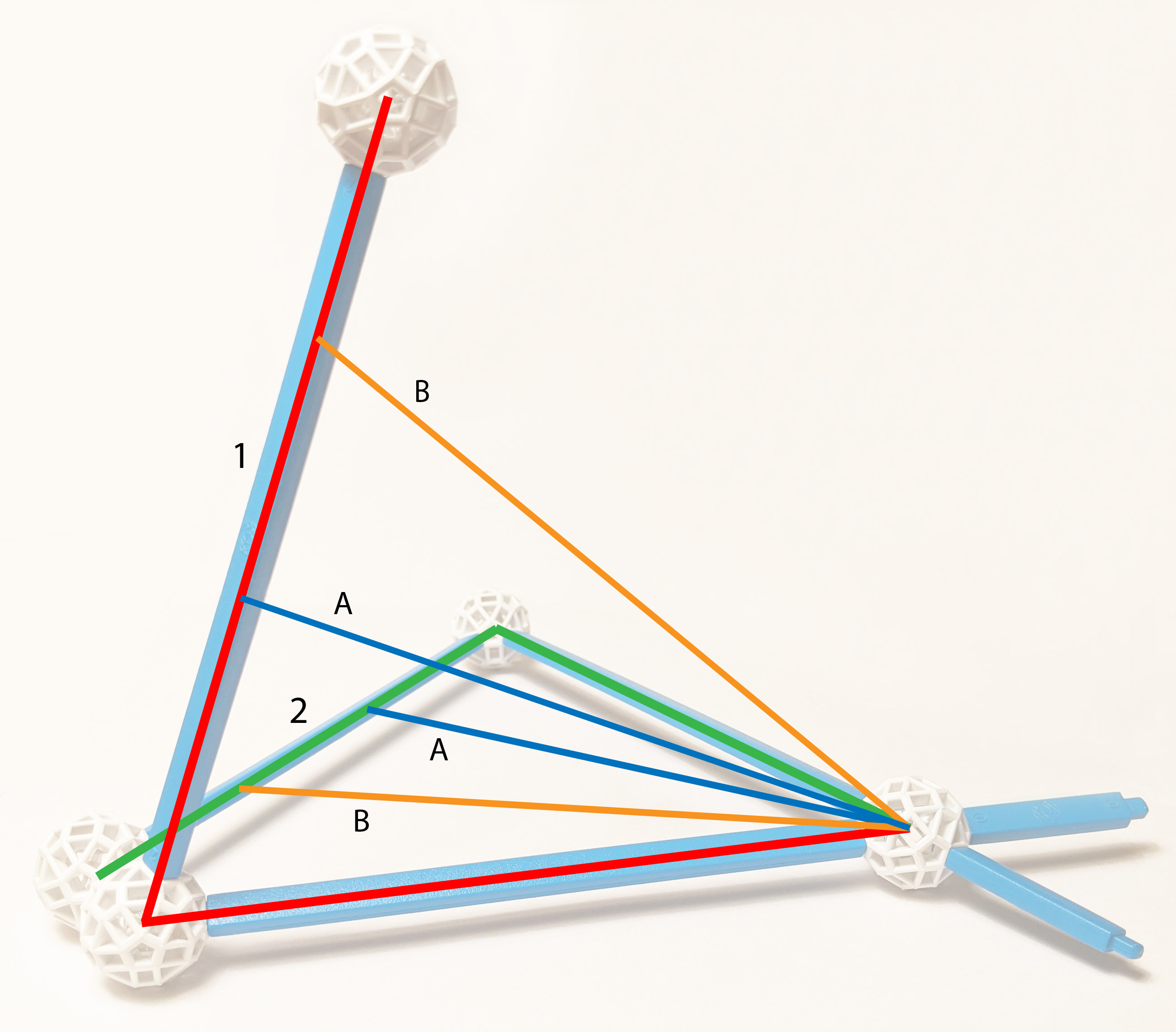}\qquad\quad\hspace*{0pt}
\includegraphics[width=.3642\textwidth]{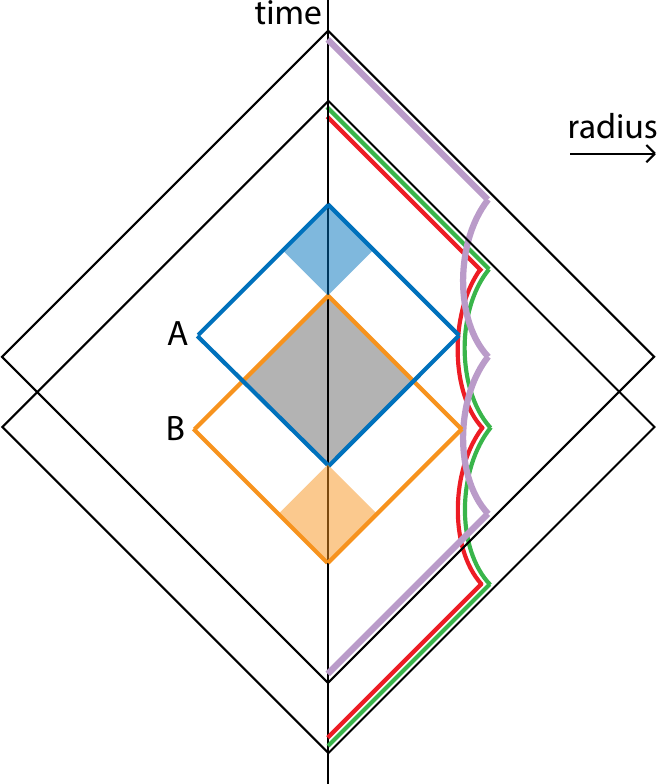}\hspace*{-263.4pt}\raisebox{.3823\textwidth}[0in][0in]{\hphantom{(b)}(a)}\hspace*{257.3pt}\vspace*{2.5pt}\\
\includegraphics[width=.4954\textwidth]{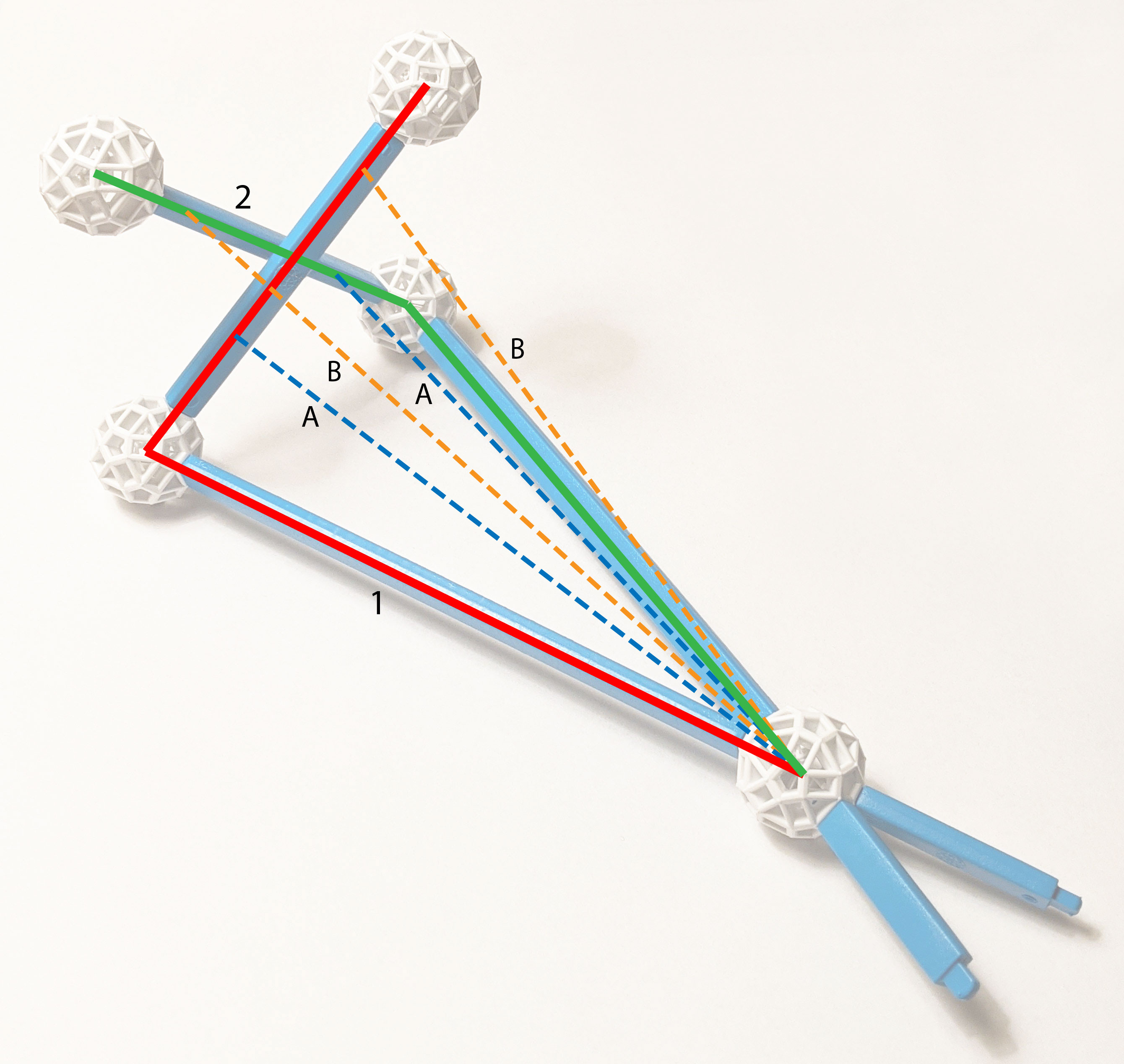}\qquad\quad\hspace*{0pt}
\raisebox{.02\textwidth}[0in][0in]{\includegraphics[width=.3642\textwidth]{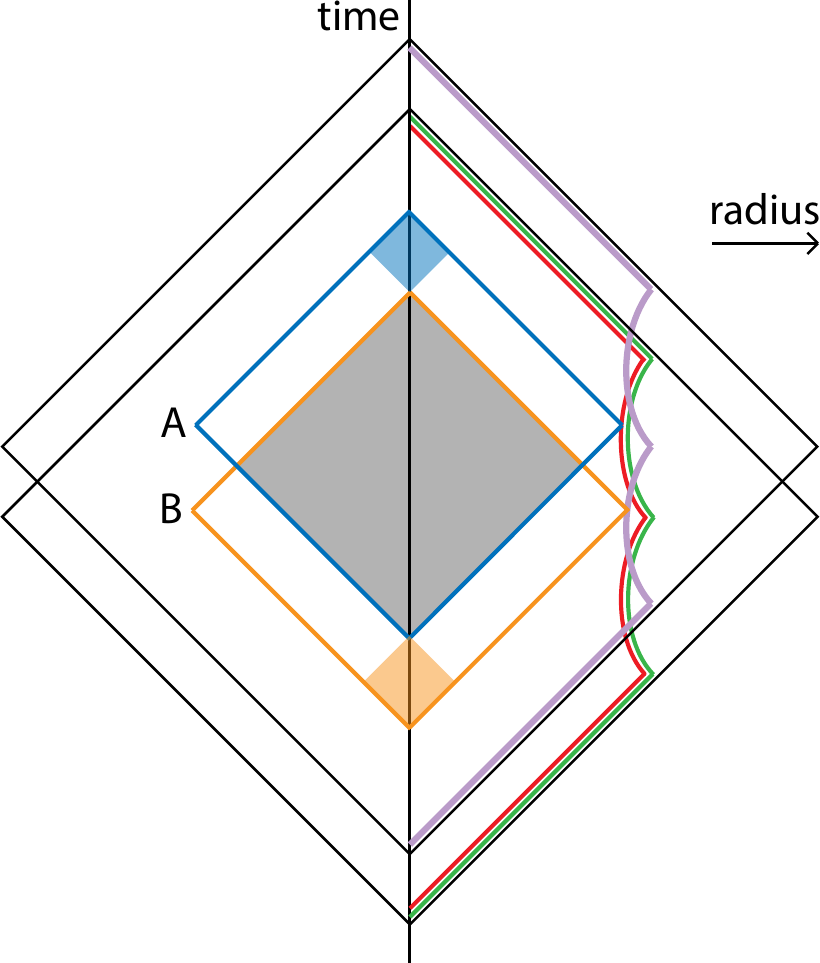}}\hspace*{-263.4pt}\raisebox{.4158\textwidth}[0in][0in]{\hphantom{(a)}(b)}\hspace*{257.3pt}\vspace*{0.pt}
\caption{Schematic diagram of interferometer correlations. Configuration (a) is described here, with configuration (b) to be interpreted analogously. \textit{Left\,}: The two arms are bent in orthogonal directions, with arm~1 (red) measuring rotations with respect to the $x$-axis and arm~2 (green) measuring rotations around the $z$-axis. The $y$-axis is defined along the radial segment of arm~1, with the bent segment going into the first quadrant of the $y$-$z$ plane. All of arm~2 sits in the first quadrant of the $x$-$y$ plane. Each pair of radial vectors denoted A (blue) or B (yellow) represents a pair of ``coincident'' points along the two arms that share the same causal diamond. \textit{Right\,}: Both arms (red and green) follow the same trajectory in a diagram of radius versus time. The whole measurement (through both arms) is delimited by a large black causal diamond. Another measurement in the same instrument, offset in time by $\tau$, is represented by a thick violet trajectory and a corresponding large black causal diamond. As currently depicted, B (yellow) is in the outgoing part of the photon path, and A (blue) is in the incoming part of the photon path, although this does not necessarily need be the case. The blue and yellow causal diamonds share an overlapping causal volume, shaded grey, where the space-time should be consistent. The small shaded light blue and light yellow causal subvolumes contribute to differential quantum geometric fluctuations between their two parent causal diamonds.\vspace{0.pt}}
\label{IFOcorr}
\end{figure}

We wish to calculate the correlated effects on $S_{2,1}(\mathscr{t})$ as a function of the offset $\tau$. To estimate these time-domain correlations, we will adopt a simplified heuristic version of the Banks-Fischler HST framework. In Fig.\,\ref{IFOcorr}, we see two causal diamonds (blue and yellow) corresponding to two positions along the space-time trajectory of the light inside the interferometer. The overlapping causal volume, shaded in grey, represents a region where the space-time should be consistent. If the two causal diamonds were entirely overlapping, there would be no relative difference between their space-times. In a relational view of constructing local space-time, measuring the same causal diamond does not give you any differential fluctuation\,--- even if there are quantum uncertainties, the measurement simply defines what space-time is for the local observer. The small shaded causal subvolumes (light blue and light yellow) represent the added information that is distinct to the respective causal diamonds (blue or yellow). Both diamonds are holographic in information, and the total quantum uncertainty, or variance, scales linearly with the size of each diamond like a random walk, but the relative fluctuation power scales with the offset between the two diamonds. The causal horizons are made out of null sequences of 2-spheres, each of which have two geometric degrees of freedom on its boundary.

We can translate this picture into a semiclassical calculational framework. Imagine the detector's space-time foliated with past and future null cones at approximately Planckian intervals, since that is the layer spacing at which we expect uncorrelated and independent quantum geometric uncertainties. In other words, we have a new blue or yellow causal diamond roughly every Planck time (with the prefactor in Eq.\,(\ref{zurekcorr}) providing the nominal normalization). This means that the light blue or light yellow region is a Planck sized causal volume, which can be thought of as the virtual 't~Hooft black hole in Fig.\,\ref{lightcone}, with the Dirac light cones in future and past directions extending out from it. We will parameterize the observer's space-time in terms of these Dirac null cones, denoted by the coordinates $\,\mathscr{T}_\pm =  t \pm {|\mathbf{r}|}/{c}$. 

\enlargethispage{1.5pt} 
As the light travels along its space-time trajectory in Fig.\,\ref{IFOcorr}, it encounters a new Planckian ``bit'' of quantum geometric uncertainty for each null cone foliation it passes through (it does not encounter any new information or uncertainty while traveling along one null cone, as this is a single coherent state), and its phase couples to this geometric uncertainty if the path is configured the right way. Again, this is only a heuristic model, and in actuality the photons are delocalized along the entire light path; the quantum space-time does not ``collapse'' until a macroscopic measurement is made. For each space-time interval of measurement (depicted by a black causal diamond), the coherent accumulation of all Planckian quantum geometric fluctuations couples to a set of boundary conditions that defines the modes of the photon field in the interferometer. For now, proceeding with our heuristic model, an approximately correct scaling behavior can be obtained by positing that for each Planck ``thickness'' Dirac light cone, the two transverse degrees of freedom on a given 2-sphere at radius~$|\mathbf{r}|$ contribute Planckian quantum geometric fluctuations, with the following covariance structure between the transverse rotational fluctuations of two position vectors $\mathbf{r}$~and~$\mathbf{r}'$ measured at times $t$~and~$t'$ respectively (the inputs are expressed in terms of the null cone coordinates $\,\smash{\mathscr{T}^{}_\pm}$ and $\,\smash{\mathscr{T}'_\pm}$, which suppress the directionality of the position vectors but provide a simple illustration of the time-correlations): 
\begin{align}
\mathrm{cov}\:\Big(\delta L^{\perp}_{z,x}(\mathscr{T}_+,\,\mathscr{T}_-),\;\delta L^{\perp}_{z,x}(\mathscr{T}'_+,\,\mathscr{T}'_-)\Big)=&\,\begin{cases}
\frac{1}{4\sqrt{2\pi}}\:\ell_{P}^{2}\;\mathbf{G}(\mathbf{r},\,\mathbf{r}')\,, & |\mathscr{T}^{}_\pm-\mathscr{T}'_\pm|<\frac{1}{2}t_{P}\ \ \:\mathrm{or}\ \ \:|\mathscr{T}^{}_\pm-\mathscr{T}'_\mp|<\frac{1}{2}t_{P}\\ \;\!0\,, & \textrm{otherwise}
\end{cases}\label{covariance}\\
\mathrm{cov}\:\Big(\delta L^{\perp}_{z}(\mathscr{T}_+,\,\mathscr{T}_-),\;\delta L^{\perp}_{x}(\mathscr{T}'_+,\,\mathscr{T}'_-)\Big)=&\:\;0\label{covarianceorthogonal}
\end{align} 
Here, we again take the modified version of $\mathbf{G}(\mathbf{r},\,\mathbf{r}')$ in Fig.\,\ref{angcorr} where $\ell=3,\,5,\,7,\,$...\,, and the subscripts $z$ and $x$ denote rotational fluctuations around those respective axes. Rotational measurements with respect to two different orthogonal axes are not correlated, although necessary to observe irreducible quantum uncertainties, as we will see shortly. Note that the Planckian covariance interval $t_P$ applies to any combination of $\,\smash{\mathscr{T}^{}_\pm}$ and $\,\smash{\mathscr{T}'_\pm}$. This implements the 't~Hooft correlations between incoming and outgoing information in a virtual Planck black hole, which asymptotically map onto the past and future parts of the Dirac light cone in Fig.\,\ref{lightcone} (there are two parity changes that cancel out, one from the antipodal identification in space-time and one from the odd harmonic modes). A normalization factor of $1/\sqrt{2\pi}$ was used based on the prefactor in Eq.\,(\ref{zurekcorr}). We could have chosen a different normalization for the Planckian variance, and instead modified the null foliation spacing to get the same overall prefactor for the macroscopic uncertainty, but this does not affect any results going forward. An additional factor of $1/4$ accounts for two factors of $1/2$ dividing the total variance: there are two transverse degrees of freedom on the 2-sphere boundary, and the total uncertainty arises from the combined effect of incoming and outgoing information. For the latter, as 't~Hooft carefully explains, the antipodal identification involving $u^\pm$ and $p^\pm$ states does not result in a double cover of the geometric degrees of freedom; in fact, it fixes the one-to-two mapping of the standard interpretation. This subtlety is not fully implemented here due to the difficulty of correctly identifying the relevant microstates including the transverse degrees of freedom, but we expect the effect to be similar, because the characteristic rate of position shift in each Planckian fluctuation is approximately one Planck length per Planck time\,--- as in, in Planck units, for each step in the Planckian random walk, the position change and the rate of change are both unity because these are null fluctuations. As noted before, specifics about the physical characteristics of the effect at this level generally do not significantly change the final frequency spectrum projected, although they can modify the intermediate level features of the time-domain correlation.

We now formulate a more general case, restoring the directionality of the position vectors. Fig.\,\ref{IFOcorr} shows that we need to calculate a correlation involving four different points in the light path. Since an interferometer output is sensitive to the optical path difference between the two arms, and we are correlating two outputs measured at different times, we consider the differential effect on the signal from two orthogonal rotations in arms 1~and~2, correlated between two pairs of points $A$~and~$B$ along the light path of each arm (where the light heuristically ``passes through at different times,'' even though the photons are actually delocalized). To avoid cluttering up our equations too much, we write down the simplified case of a covariance on a single Planckian null boundary layer, such that the condition of the point pairs $A$~and~$B$ being on the same Dirac null cone foliation in Eq.\,(\ref{covariance}) is already satisfied (accounting for the different spatial positions $\mathbf{r}$ and laboratory times $t$ where the light path intersects the Planckian null boundary layer):\vspace{.pt}
\begin{equation}
\mathrm{cov}\:\Big(\delta L_z^A (\mathbf{r}_2^A)- \delta L_x^A (\mathbf{r}_1^A),\;\delta L_z^B (\mathbf{r}_2^B)- \delta L_x^B (\mathbf{r}_1^B)\Big)=\,\mathrm{cov}\:\Big(\delta L_z^A (\mathbf{r}_2^A),\;\delta L_z^B (\mathbf{r}_2^B)\Big)+\,\mathrm{cov}\:\Big(\delta L_x^A (\mathbf{r}_1^A),\;\delta L_x^B (\mathbf{r}_1^B)\Big)\vspace{.pt}
\label{darmcov}\end{equation}
We see that there is no covariance between rotational fluctuations in orthogonal directions, so the cross-terms drop out. We can also see, however, that measuring two incompatible observables along orthogonal rotational axes results in an irreducible quantum indeterminacy, by writing down the further simplified case where $A=B$:
\begin{align}
\,\mathrm{cov}\:\Big(\delta L_z^A (\mathbf{r}_2^A),\;\delta L_z^B (\mathbf{r}_2^B)\Big)+\,\mathrm{cov}\:\Big(\delta L_x^A (\mathbf{r}_1^A),\;\delta L_x^B (\mathbf{r}_1^B)\Big) &= \,\delta L_z^\perp (\mathbf{r}_2)^2 \, +\, \delta L_x^\perp (\mathbf{r}_1)^2 \,\gtrsim\: 2 \left|\delta L_z^\perp (\mathbf{r}_2)\right|\left|\delta L_x^\perp (\mathbf{r}_1)\right|\,\label{covuncertainty}\\
&\approx \,\frac{1}{2\sqrt{2\pi}}\;\ell_{P}^{2}\,\left|\,\sum_{\ell=3}^\mathrm{odd}\:\frac{1}{4\pi}\:\frac{2\ell+1}{\ell^2+\ell+1} \  P_\ell\,(\:\!\hat{\mathbf{r}}_2 \cdot \hat{\mathbf{r}}_1)\,\right|\label{covangular}
\end{align}
Here, $\hat{\mathbf{r}}_1$ and $\hat{\mathbf{r}}_2$ denote unit vectors in the directions of $\mathbf{r}_1$ and $\mathbf{r}_2$ respectively. We may find the angular correlator in Eq.\,(\ref{covangular}) odd since it pertains to two arms measuring orthogonal rotational degrees of freedom, which we claimed are not correlated. However, the idea here is that the amount of incompatibility between these observables depends on the amount of correlation there is for each of those degrees of freedom respectively. For example, $\delta L_z^\perp (\mathbf{r}_1)$ and $\delta L_z^\perp (\mathbf{r}_2)$, the $z$ rotations at $\mathbf{r}_1$ and $\mathbf{r}_2$ respectively, 
are correlated by a factor $P_\ell\,(\:\!\hat{\mathbf{r}}_1 \cdot \hat{\mathbf{r}}_2)$, and since measuring $\delta L_x^\perp (\mathbf{r}_1)$ and $\delta L_z^\perp (\mathbf{r}_1)$, the $x$ and $z$ rotations at $\mathbf{r}_1$ respectively, forms an irreducible uncertainty between them, the amount of indeterminacy we encounter between $\delta L_x^\perp (\mathbf{r}_1)$ and $\delta L_z^\perp (\mathbf{r}_2)$ is proportional to this angular correlator. The angular correlations become considerably more complicated when there are four points, as in $A\neq B$. A simple way to visualize this would be to consider one more angular correlator for each $\ell$ mode, $P_\ell\,(\:\!\hat{\mathbf{r}}^A_1 \cdot \hat{\mathbf{r}}^B_1) = P_\ell\,(\:\!\hat{\mathbf{r}}^A_2 \cdot \hat{\mathbf{r}}^B_2)$, although obviously this angular correlation is not independent from the one in Eq.\,(\ref{covangular}). We will omit for the purposes of this manuscript this level of detail in the implementation of the angular correlators. Again, as will soon become clear in our results, the final shape and magnitude of the projected frequency spectrum are quite stable to differences in such details.

We are finally ready to evaluate the time-domain differential arm length signals. First, we write a semiclassical expression coupling $S_{2,1}(\mathscr{t})$ to the Planckian transverse fluctuations along the light path:\vspace{.pt}
\begin{equation}
S_{2,1}(\mathscr{t}) = \int_{\mathscr{t}-\mathcal{T}}^{\mathscr{t}}\left[\dot{\mathbf{r}}_{2,1}(t)+{\textstyle\sum}_\pm\dot{L}^{\pm}_{z,x}\big(t,\,\mathbf{r}_{2,1}(t)\big)\,\boldsymbol{\hat{\theta}}_{z,x}\right]\cdot\frac{\dot{\mathbf{r}}_{2,1}(t)}{c}\:\mathrm{d}t\vspace{.pt} 
\label{singlepath}\end{equation}
Here, $\mathcal{T}=2L/c$ is the light travel time inside the interferometer, and $\smash{\boldsymbol{\hat{\theta}}_{z,x}}$ denotes a unit vector in the transverse direction corresponding to a rotation around the $z$ or $x$ axis. The pair of indices (2,\,1) and ($z$,\,$x$) should be matched, as each arm is only sensitive to its respective rotational axis. The first term in the square brackets is simply the light propagating along its path, and the second holographic space-time fluctuation term $\smash{\dot{L}^{\pm}_{z,x}(t,\,\mathbf{r}_{2,1}(t))}$ is defined as:\vspace{.pt}
\begin{equation}
\dot{L}^{\pm}_{z,x}\big(t,\,\mathbf{r}_{2,1}(t)\big)\equiv\,\frac{\delta{L}^{\perp}_{z,x}\big(t,\,\mathbf{r}_{2,1}(t)\big)}{t_P}\:\frac{\mathrm{d}\mathscr{T}_\pm(t,\,\mathbf{r}_{2,1}(t)\big)}{\mathrm{d}t}\vspace{.pt}
\label{fluctuate}\end{equation}
This should be interpreted as an expression of ``fluctuation rate'' in the context of a Planckian random walk, where the step size $\delta \mathscr{T}_\pm = t_P$ is used as the first denominator. Note that we have two terms of $\smash{\dot{L}^{\pm}_{z,x}(t,\,\mathbf{r}_{2,1}(t))}$ because for each point $(t,\,\mathbf{r}_{2,1}(t))$ we have both past and future light cones passing through it, denoted by the $\pm$ symbols. The first fraction here is the familiar transverse rotational fluctuation from Eqs.\,(\ref{covariance}\,--\,\ref{covangular}), just divided by $t_P$ to express the fact that this Planckian jitter happens within a null cone foliation layer of Planck time thickness. The second fraction represents the fact that the relationship between $\mathscr{T}_\pm$ and $t$ is nonlinear; since the light has nonzero inward or outward radial velocity throughout its trajectory, one Planck time in laboratory time does not equal one Planck time in the null cone coordinates. The light path sometimes intersects many layers of null cone foliations quickly, and sometimes it runs parallel to them, not passing through any at all. We now rewrite $S_{2,1}(\mathscr{t})$ in terms of null cone coordinates:\vspace{-.5pt}
\begin{equation}
\!\!S_{2,1}(\mathscr{t}) = \,c\,\mathcal{T}+\int_{\mathscr{t}-\mathcal{T}}^{\mathscr{t}}\!\frac{\delta{L}^{\perp}_{z,x}\big(\mathscr{T}_+,\,\mathbf{r}_{2,1}(\mathscr{T}_+)\big)}{t_P}\,\boldsymbol{\hat{\theta}}_{z,x}\cdot\frac{\dot{\mathbf{r}}_{2,1}(\mathscr{T}_+)}{c}\:\mathrm{d}\mathscr{T}_+ +\int_{\mathscr{t}-\mathcal{T}}^{\mathscr{t}}\!\frac{\delta{L}^{\perp}_{z,x}\big(\mathscr{T}_-,\,\mathbf{r}_{2,1}(\mathscr{T}_-)\big)}{t_P}\,\boldsymbol{\hat{\theta}}_{z,x}\cdot\frac{\dot{\mathbf{r}}_{2,1}(\mathscr{T}_-)}{c}\:\mathrm{d}\mathscr{T}_-\vspace{-.5pt}\label{singlepathnulltime}
\end{equation}

This can be combined with Eq.\,(\ref{DARMcorr}) to form the full expression for the time-domain DARM correlation function, where we will reparameterize $\mathbf{r}_{2,1}(t)$ to set $\mathscr{t}=\mathcal{T}$ without loss of generality (as in, the light was injected at $t=0$):\vspace{-.5pt}
\begin{align}
&\mathrm{DARM\;Corr}\,(\tau)\,= \:\frac{1}{4}\;\sum_{\zeta,\,\eta}^{\pm}\sum_{\alpha,\,\beta}^{(2,z)(1,x)}(-1)^{\alpha+\beta}\;\Bigg\langle \int_{0}^{\mathcal{T}}\frac{\delta{L}^{\perp}_{\alpha}\big(\mathscr{T}_\zeta,\,\mathbf{r}_{\alpha}(\mathscr{T}_\zeta)\big)}{t_P}\,\boldsymbol{\hat{\theta}}_{\alpha}\cdot\frac{\dot{\mathbf{r}}_{\alpha}(\mathscr{T}_\zeta)}{c}\:\mathrm{d}\mathscr{T}_\zeta\nonumber \\
&\qquad\qquad\qquad\qquad\qquad\qquad\qquad\qquad\qquad\hspace*{.583em}\times \int_{\tau}^{\tau+\mathcal{T}}\frac{\delta{L}^{\perp}_{\beta}\big(\mathscr{T}'_\eta,\,\mathbf{r}_{\beta}(\mathscr{T}'_\eta)\big)}{t_P}\,\boldsymbol{\hat{\theta}}_{\beta}\cdot\frac{\dot{\mathbf{r}}_{\beta}(\mathscr{T}'_\eta)}{c}\:\mathrm{d}\mathscr{T}'_\eta\Bigg\rangle\label{DARMcorrlong}\\
&\;\,= \: \frac{1}{4\:\!t_P^2}\;\sum_{\zeta,\,\eta}^{\pm}\sum_{\alpha}^{(2,z)(1,x)} \int_{0}^{\mathcal{T}}\boldsymbol{\hat{\theta}}_{\alpha}\cdot\frac{\dot{\mathbf{r}}_{\alpha}(\mathscr{T}_\zeta)}{c}\:\mathrm{d}\mathscr{T}_\zeta \;\int_{\tau}^{\tau+\mathcal{T}}\boldsymbol{\hat{\theta}}_{\alpha}\cdot\frac{\dot{\mathbf{r}}_{\alpha}(\mathscr{T}'_\eta)}{c}\:\mathrm{d}\mathscr{T}'_\eta \:\; \Big\langle\delta{L}^{\perp}_{\alpha}\big(\mathscr{T}_\zeta,\,\mathbf{r}_{\alpha}(\mathscr{T}_\zeta)\big)\; \delta{L}^{\perp}_{\alpha}\big(\mathscr{T}'_\eta,\,\mathbf{r}_{\alpha}(\mathscr{T}'_\eta)\big)\Big\rangle\:\label{DARMcorrcollect}
\end{align}
Here, we have used Eq.\,(\ref{darmcov}) to eliminate the cross-terms evaluating correlations between the two arms, as they measure orthogonal rotational degrees of freedom. In fact, the quantity in the angle brackets in Eq.\,(\ref{DARMcorrcollect}), summed over the two arms $\alpha=(2,z)\ \mathrm{or}\ (1,x)$, is precisely the quantity on the right hand side of Eq.\,(\ref{darmcov}). Since this covariance is only nonzero within a Planckian foliation layer, we can use the term to eliminate one of the integrals. Its contribution will be a value of \:\!$\sim\:\!\!\ell_P^2$ over an microscopic interval $t_P$, for an integral of \:\!$\sim\:\!\! \ell_P^2\,t_P$. The remaining integral will run over the system scale \:\!$\sim\:\!\!\mathcal{T}$, so combined with the prefactor \:\!$\sim\:\!\!1/t_P^2$, the entire expression should evaluate to a value \:\!$\sim\:\!\! c\,\ell_P\,\mathcal{T}$, just as we expected. Unfortunately, due to the complexity, most of the calculation will be done numerically.

\begin{figure}[t]
\centering
\vspace*{-3.pt}\hspace*{-23pt}\includegraphics[height=.23\textheight]{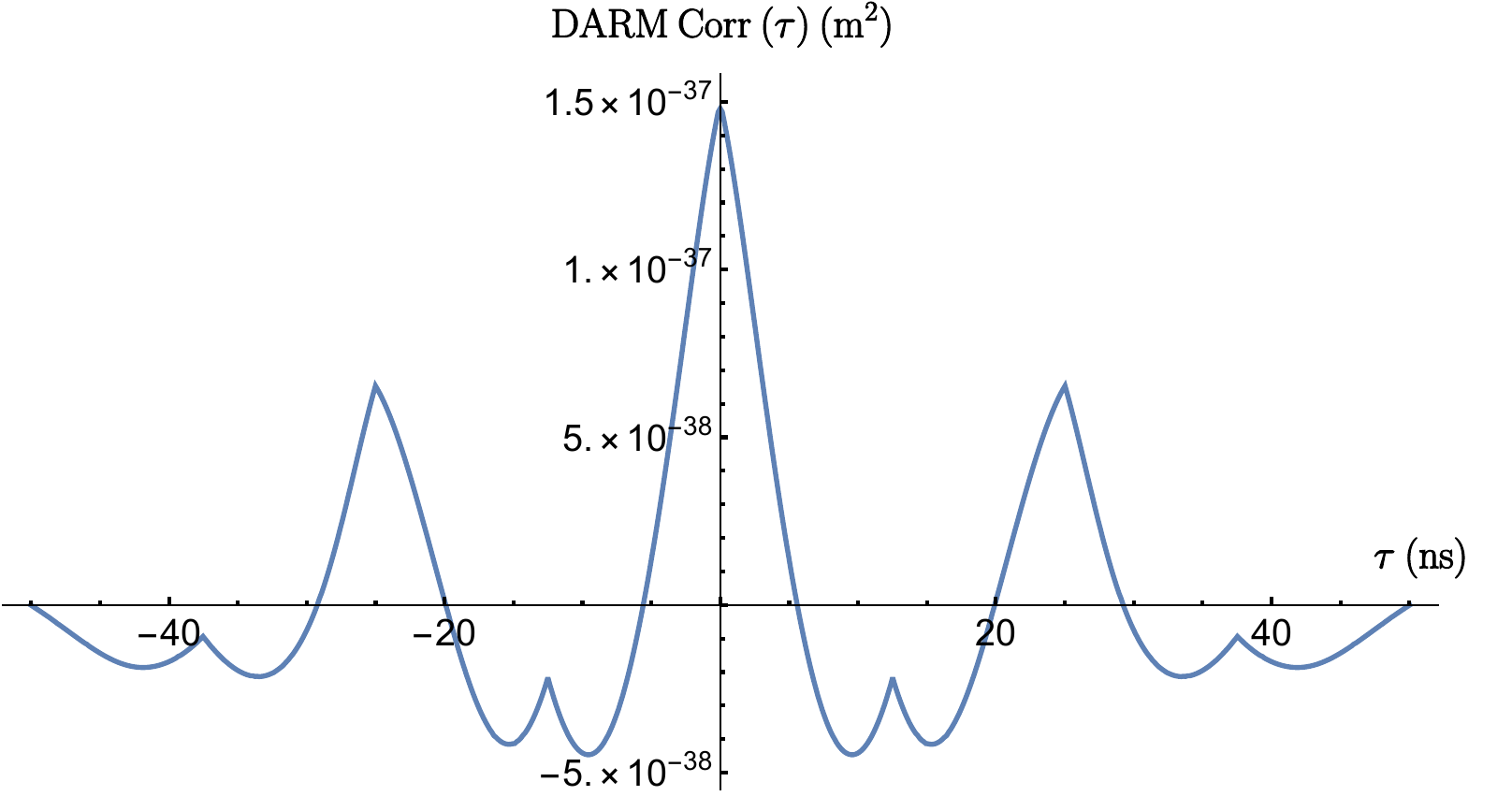}\hfill
\includegraphics[height=.23\textheight]{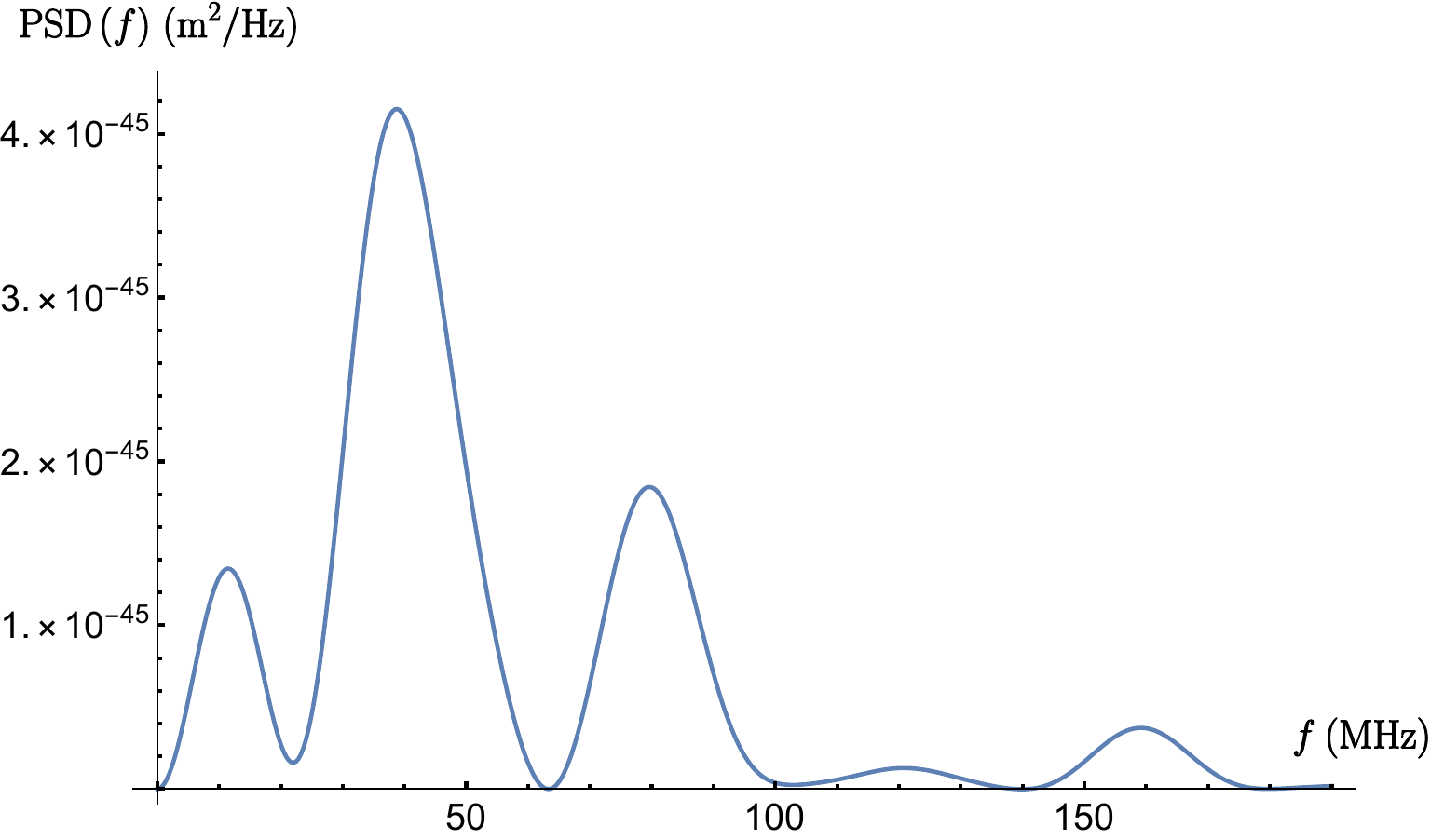}\hspace*{-23pt} \\
\vspace{-134pt}\hfill\textbf{(a)}\qquad\qquad\qquad\qquad\qquad\qquad\qquad\qquad\qquad\qquad\qquad\quad\hspace*{26.9pt}\textbf{(a)}\vspace{122.5pt} \\ \vspace*{9pt}
\hspace*{-23pt}\includegraphics[height=.23\textheight]{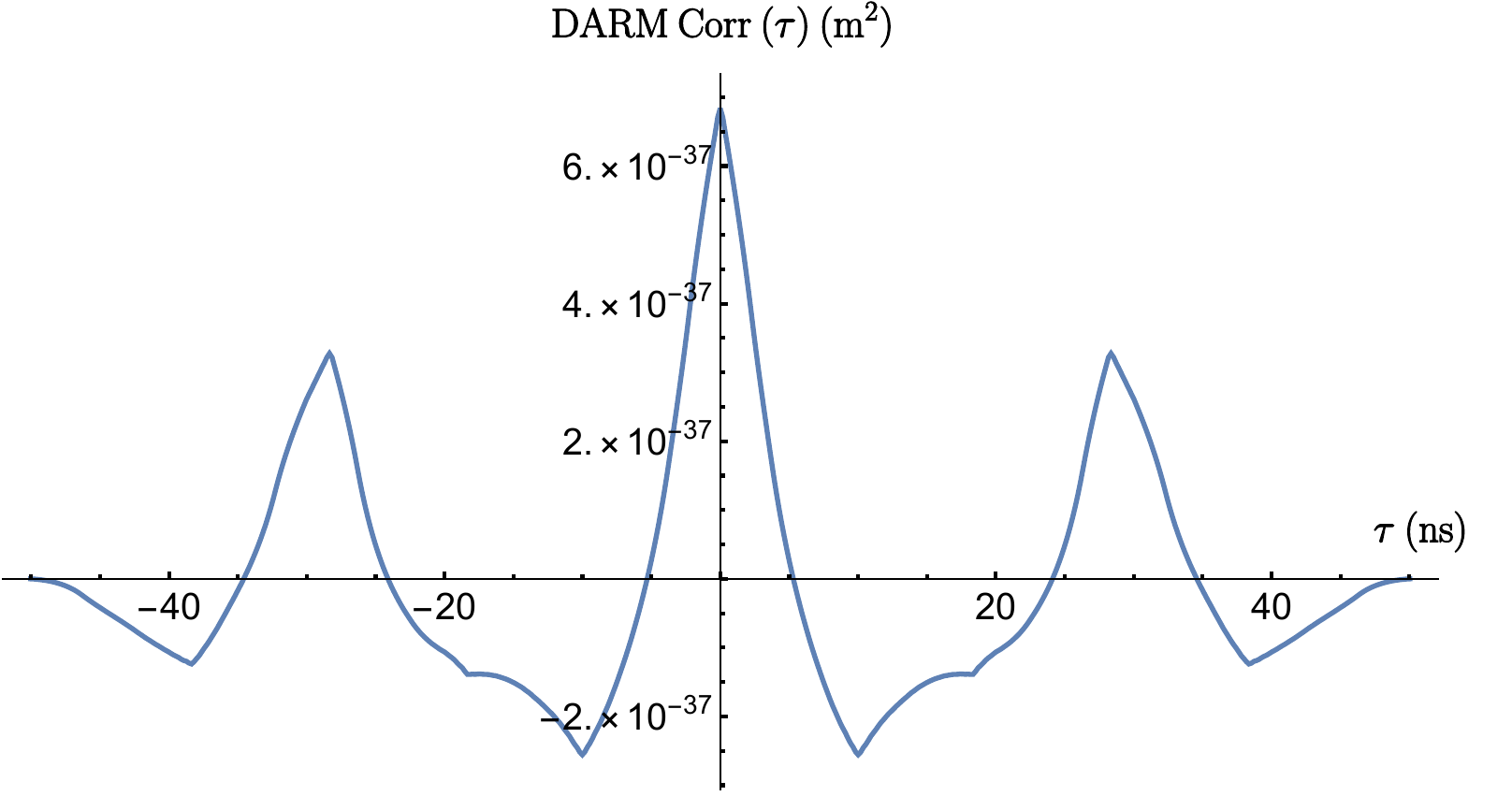}\hfill
\includegraphics[height=.23\textheight]{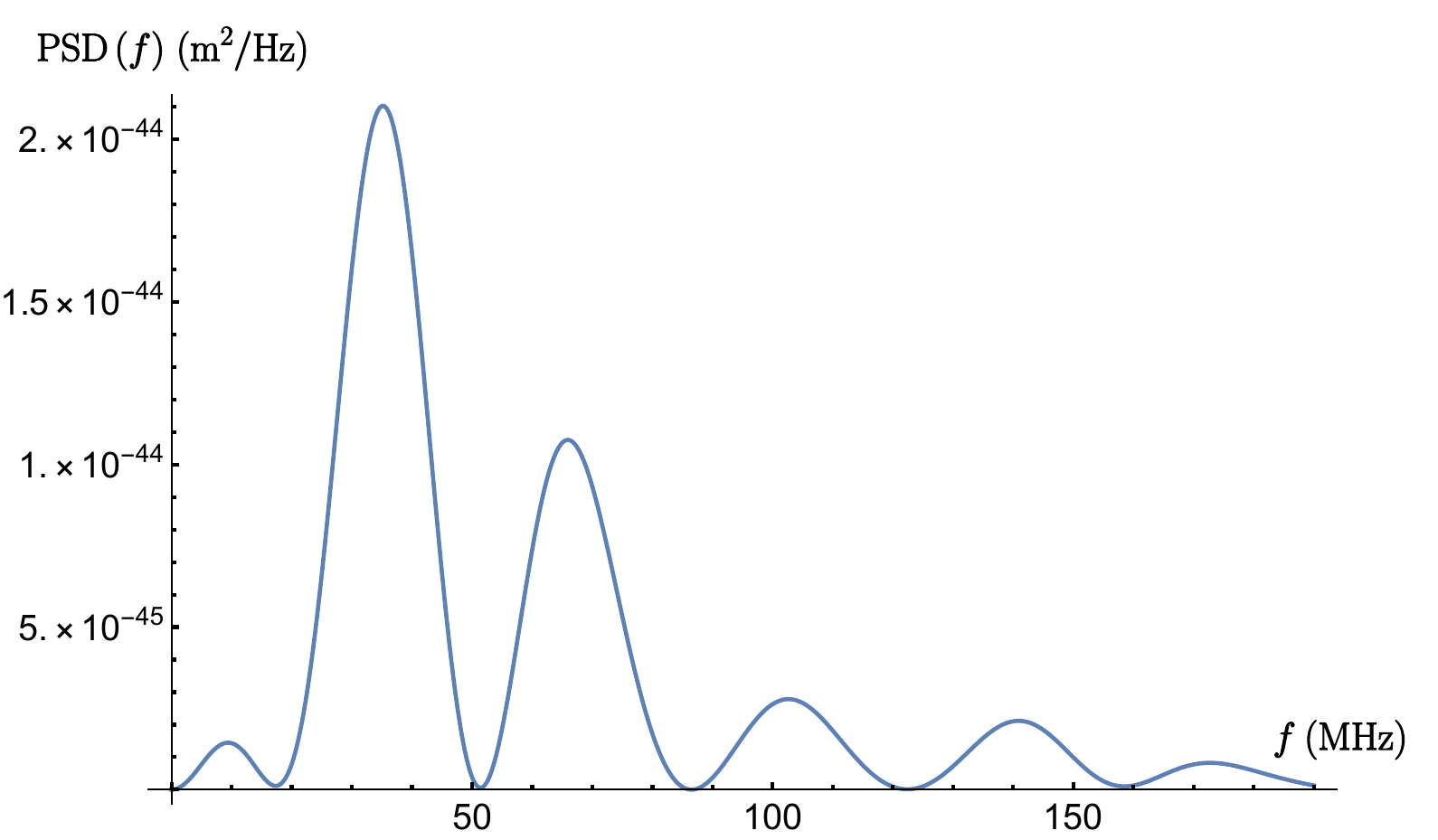}\hspace*{-23pt} \\
\vspace{-134pt}\hfill\textbf{(b)}\qquad\qquad\qquad\qquad\qquad\qquad\qquad\qquad\qquad\qquad\qquad\quad\hspace*{26.1pt}\textbf{(b)}\vspace{122.5pt}\vspace*{-.5pt}
\caption{The time-domain DARM correlation and frequency-domain power spectral density (PSD) for each of the proposed geometric configurations (a) and (b), estimated for interferometers of size $L=7.5\,$m. The characteristic scales of the holographic space-time fluctuations for this system size are $\mathrm{DARM\;Corr}\,(\tau)\sim\ell_P L \approx 1.2\times10^{-34}\,\mathrm{m}^2$ and $\tau\sim L/c=25\,\mathrm{ns}$ in the time domain, and $\mathrm{PSD}\,(f)\sim t_P L^2\approx 3\times10^{-42}\,\mathrm{m}^2/\mathrm{Hz}$ and $f\sim c/L\approx40\,\mathrm{MHz}$ in the frequency domain. We see that all estimated spectra are 2\,--\,3 orders of magnitude smaller than the characteristic scales, due to the instrument's partial coupling to the effect, the angular correlation functions, and the division of fluctuation power among many different modes.\vspace{-.0pt}}
\label{spectra}
\end{figure}

A few notes about the calculation: The function $\mathbf{r}_{2,1}(\mathscr{T}_\pm)$ is a parameterization of the light trajectory inside the interferometer in terms of the null cone coordinates $\mathscr{T}_\pm$. The inner product $\smash{\boldsymbol{\hat{\theta}}_{z,x}\cdot\dot{\mathbf{r}}_{2,1}(\mathscr{T})/c}$, representing the projection of the transverse rotational fluctuations onto a direction tangent to the light path, can take positive or negative values depending on whether the light is in the outgoing or incoming portion of the trajectory. This angular factor is applied on top of the four Legendre polynomial angular correlators discussed above, between two points within each arm and between the two arms (the mode summation in Eq.\,(\ref{covangular}) is only applied once, of course). The $(\zeta,\,\eta)=(\pm,\,\pm)$ integrals, representing correlations on null cone foliations in the same direction, and the $(\zeta,\,\eta)=(\pm,\,\mp)$ integrals, representing correlations between past and future null cones, obviously do not all run the entire range, as the covariance is only nonzero when there is overlap in the null cone times. In practice, the integral in Eq.\,(\ref{DARMcorrcollect}) is evaluated in 16 segments for an interferometer that has bent and radial portions whose lengths are identical between the two arms, accounting for the fact that we only need to evaluate it for $\tau>0$ since the DARM correlation is symmetric.

\enlargethispage{3.5pt}
Finally, we evaluate the power spectral density in the frequency domain. Using the engineering convention, where the fluctuation power in the negative frequencies is folded into the positive frequencies, the PSD is:\vspace{-1.75pt}
\begin{equation}\label{PSD}
\mathrm{PSD}\,(f) \equiv \,4\int_{0}^{\infty}\mathrm{DARM\;Corr}\,(\tau)\:\cos\left(2\pi f\tau\right)\,\mathrm{d}\tau\vspace{-2pt}
\end{equation}

The time-domain and frequency-domain results are plotted in Fig.\,\ref{spectra} for both configurations (a) and (b) presented in Fig.\,\ref{configs}. We see that the overall time-domain structures are similar for the two interferometer geometries, and that the peak frequencies and the distribution of power at those frequencies only differ by about 10\,--\,20\% despite the very different designs. Using configuration (a) as an example, the general time-domain structure owes to the three scales at play here: the round-trip distance of the whole arm, and the one-way and round-trip distances of the bent segment. The former decides the overall extent of the correlation, and the latter two scales determine the locations of the sharp discontinuities in the slope, as the correlation shifts between outgoing and incoming light, and between null cone foliations in the same direction versus crossings of past\,/\,future null cones. The same structure mostly holds for configuration (b), although the correlation involves more scales because the bent and radial segments are different in length. One additional feature of note for its time-domain plot is that some of the discontinuities in the slope are due to the points midway through the bent segments where the two arms intersect orthogonally. As explained before, the logarithmic divergence at such a point is well-controlled in our model, but this point is still a strong maximum in the angular correlation, and thereby leads to extrema in the time-domain features.

\begin{figure}[t]
\centering
\vspace{-1.5pt}\includegraphics[height=.2925\textheight]{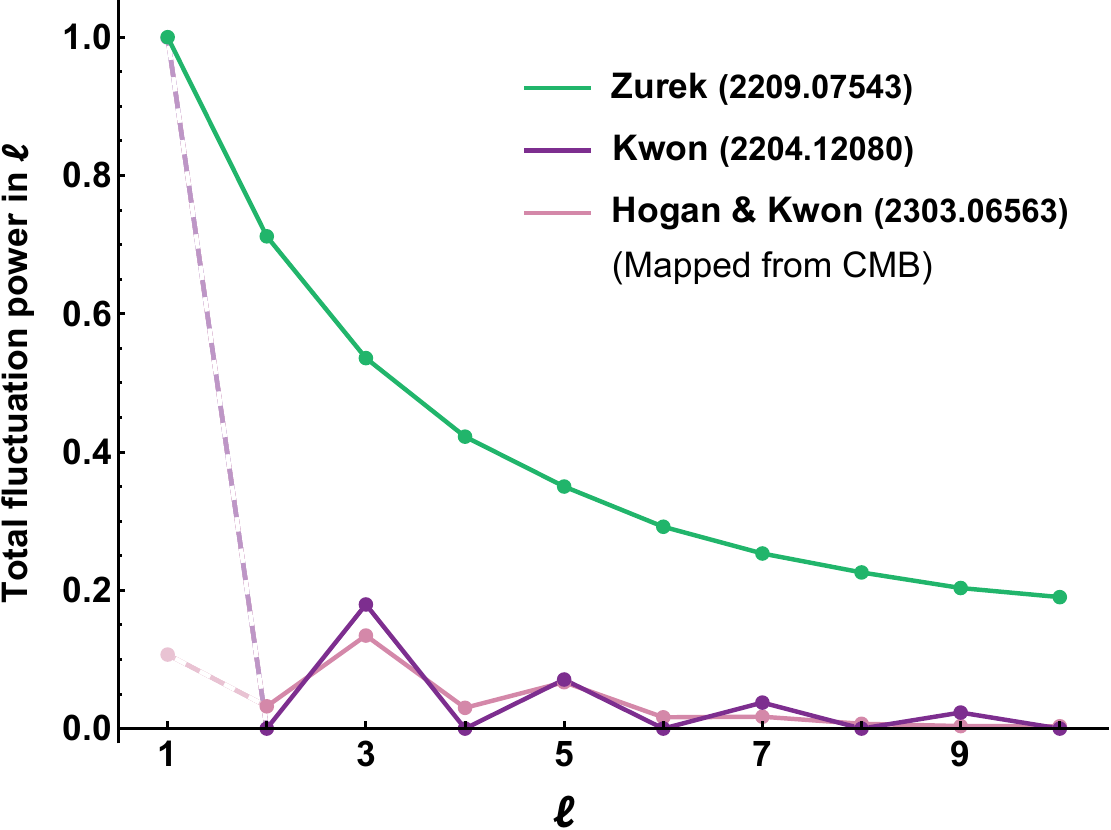}\vspace{-1.5pt}
\caption{The distribution of observable fluctuation power across different $\ell$ modes (summed over the $m$ modes). The ``Kwon (2204.12080)'' spectrum, the prediction in this work matching the correlation functions in Fig.\,\ref{spectra}, and the ``Zurek (2209.07543)'' spectrum, from Fig.\,4 of Ref.\,\cite{LiPRD}, are normalized to their respective $\ell=1$ mode powers. The ``Hogan \& Kwon (2303.06563)'' spectrum, from the Appendix of our follow-up work Ref.\,\cite{hogan2023angular} where apparent holographic symmetries in the CMB were mapped onto flat space-time, is normalized to the same observable fluctuation power as the ``Kwon (2204.12080)'' spectrum. The ``Zurek (2209.07543)'' spectrum scales approximately as \:\!$\sim\:\!\!1/\ell$, whereas the ``Kwon (2204.12080)'' and ``Hogan \& Kwon (2303.06563)'' spectra scale roughly as \:\!$\sim\:\!\!1/\ell^2$, with a distinct dominance of odd modes over even modes. The faded values at $\ell=1$ are not physically meaningful, as this mode is unobservable in the CMB and has a suppressed coupling to observables in flat space-time.\vspace{-.5pt}}
\label{angspec}
\end{figure}

\enlargethispage{4pt} 
These correlation functions overall have a different distribution of observable fluctuation power over the $\ell$ modes compared to the unadjusted Verlinde-Zurek decomposition in Eq.\,(\ref{zurekcorr}). As Fig.\,\ref{angspec} shows, an angular power spectrum computed by Zurek's team\,\cite{LiPRD} based on a straightforward interpretation of the Verlinde-Zurek model as radial scalar potential fluctuations shows an approximately \:\!$\sim\:\!\!1/\ell$ scaling, whereas the correlation functions predicted in Fig.\,\ref{spectra} have signal powers whose decomposition into different angular scales follows roughly a \:\!$\sim\:\!\!1/\ell^2$ envelope. Furthermore, our predicted spectrum only contains odd modes by design, whereas the Verlinde-Zurek model contains all modes, odd and even. As discussed earlier, most of the differences originate from our considerations about the measurement problem in quantum space-time. The antisymmetric mode selection is a result of taking 't~Hooft's coherent quantum states on causal horizons\,--- and the antipodal entanglement with a sign flip\,--- seriously at face value, rather than the Verlinde-Zurek interpretation that uses effective field theory to map what should be quantum observables and operators onto definite metric fluctuations and generates the angular correlations via a Green's function instead of an $S$-matrix. The additional factor of $1/\ell$ in the angular power spectrum also arises from positing a signature that actually originates from coherent quantum states on causal boundaries instead of definite scalar metric fluctuations. In such a case, modeling the signal requires careful consideration of the observer and the quantum measurement, and of constructing space-time observables without a background. Because we predict a signal only in measurements that couple to differential fluctuations between two macroscopically separated causal diamonds, and because in our experiment this differential fluctuation scales with the extent of transverse photon propagation within a single measurement, there is an additional linear factor of angular scale in the predicted \textit{observable} fluctuation power.

Fig.\,\ref{angspec} shows another spectrum for comparison, labeled ``Mapped from CMB,'' from the Appendix of our follow-up work Ref.\,\cite{hogan2023angular}. This spectrum appears to closely follow the angular power spectrum calculated in this manuscript. This is surprising, because our prediction was calculated without any inputs, and this ``Mapped from CMB'' spectrum was generated from empirical data after our prediction was made. In generating this data-driven spectrum, we took a map of cosmic microwave background anisotropy, and using basic assumptions about a causally consistent theory of inflation where the primordial fluctuations are set on causal boundaries (instead of the acausal infinite plane wave modes used in standard inflation), found an ansatz for the angular correlation function that matches the data at large angular scales (or low $\ell$) where the primordial quantum effects are expected to be dominant\,\cite{Hogan_2023}. We then mapped this ansatz onto flat space-time by taking out the expansionary history from the model, using the fact that flat space-time and inflation have the same conformal causal structure\,\cite{hogan2023angular}.

\enlargethispage{12.5pt} 
The three spectra being compared in Fig.\,\ref{angspec} are not exactly equivalent to one another, as the prediction in this work is for rotational fluctuations and the other two spectra\,--- the prediction by Zurek's team\,\cite{LiPRD} and the spectrum mapped from CMB data\,\cite{hogan2023angular}\,--- are for radial fluctuations. Still, the comparison provides some meaningful insights. While we have argued against radial fluctations being straightforwardly measurable in a simple Michelson interferometer as proposed by Verlinde-Zurek, we have also argued that if the radial fluctuations were in fact being measured in an experiment, a prediction that handles correctly the transverse quantization of the coherent quantum states on causal horizons would show a \:\!$\sim\:\!\!1/\ell^2$ scaling for observable radial fluctations, just like our prediction for rotational fluctuations (this scaling also satisfies scale invariance, which is not correctly handled by conformal field theory in this context\,\cite{hogan2023angular}). For the CMB correlations, on the other hand, the radial fluctuations are in fact straightforwardly measurable because they are ``frozen in'' as they cross the inflationary horizon; the measurement problem in quantum space-time is automatically taken care of, because the quantum uncertainties ``collapse'' to become classical perturbations at the end of inflation. It is significant that the measured CMB spectrum shows the same \:\!$\sim\:\!\!1/\ell^2$ scaling and the dominance of odd modes over even modes that we expect from purely theoretical considerations.

The modeled signals in Fig.\,\ref{spectra} are within reach of the interferometer being commissioned at Cardiff University\,\cite{grote2020,QIPublic}, despite being 2\,--\,3 orders of magnitude smaller than the characteristic scales corresponding to a simple Planckian random walk. Using configuration (b) as the nominal 3D setup, an interferometer of size $L=7.5\,$m circulating 10\,kW of laser power and deploying 6\,dB of quantum optical squeezing is expected to reach $5\:\!\sigma$ sensitivity in about 1 year of integration time. However, this does not offer any ability to map out the spectrum in the case of a detection. Thus, an upgrade is being developed in which a series of optical filters are deployed to separate out the signal photons from the carrier photons. The idea is that a detected signal at \:\!$\sim\:\!\!35\,$MHz puts a few signal photons into a sideband that is offset from the carrier photons by \:\!$\sim\:\!\!35\,$MHz, and if we can read out only these sideband signal photons with single photon detectors while suppressing the carrier light by 22\,--\,25 orders of magnitude, we can eliminate the dominant shot noise entirely, and be limited only by the subleading thermal noise. This novel readout scheme, which is also being developed by the GQuEST team\,\cite{GQuESTPublic,McCuller:2022hum,vermeulen2024}, was originally proposed in the Quantum Noise Working Group of the LIGO-Virgo-KAGRA Scientific Collaboration a few years back\,\cite{InternalLIGO} and found to confer no benefit in LIGO, but recently further studied as a powerful technique for other types of detectors\,\cite{McCuller:2022hum,vermeulen2024,yu2023photon}.

The predicted signal for an experiment using single photon readout is expected to be broadly similar to the one presented here, but involves some subtleties that require further study. The cross-correlation between two colocated interferometers can no longer be done the classical way, because the output no longer linearly measures the signal photon field. The measured photon count is quadratic in signal\,--- you can only measure ``collapsed'' signal photons, or signal power, and you lose all the phase information in the amplitude. Furthermore, even if the signal is coherent between the two interferometers, the photon emission events are random and not necessarily correlated in time or frequency. Thus, to cross-compare the signal fields before the photons ``collapse,'' the two interferometers must be fed with phase-locked lasers, and the two outputs interfered via a beamsplitter in a way such that one side constructively gets all of the coherent signal and the other destructively gets none\,\cite{McCuller:2022hum}. Since the two sides get the same incoherent shot and thermal noise, we can subtract one from the other, with the uncertainty scaling down as $1/\sqrt{n}$, where $n$ is the number of independent measurements. This recombination has the effect of erasing the ``which path'' information and entangling the photons from the two outputs, each representing a quantum state delocalized across both arms of its respective interferometer. The coupling of our quantum space-time signal to such an entangled state will be the subject of future work.\vspace{-6.5pt}

\enlargethispage{6pt} 
\section*{\NoCaseChange{Conclusion}}

In light of recent theoretical models of space-time fluctuations on holographic causal boundaries\,\cite{Verlinde_2021,Verlinde:2019ade,Banks_2021,Verlinde:2022hhs,banks2023fluc,Hooft:2016itl,*Hooft:2016cpw,*Hooft2016_,Betzios_2016,tHooft2018,Hooft:2018gtw,*Hooft:2018syc,*hooft2019,Gaddam_2022} that add to Hogan's original ``Holographic Noise'' hypothesis\,\cite{Hogan:2008a}, a phenomenological model was presented that describes how the current state of foundational insights might connect to observables measured in a specific instrument, along with a proposed design geometry for potential detection. We relied on a conceptual picture of quantum space-time where the background independence of general relativity is intricately linked to the rejection of local realism in quantum mechanics. In this picture, locality and the fabric of space-time are both built relationally. Causal horizons are where space-time states decohere for an observer, forming a quantum-classical boundary where an indeterminate quantum state becomes a concrete measured reality. Based on this conjecture, we built an interferometer design optimized for a robust signature, and calculated an approximate projected spectrum. Because we take seriously the distortion of causal structure from quantized gravitational states on all scales\,\cite{Hollands_2004,Stamp_2015,krishogan,mackewicz2024,Hooft:2016itl,*Hooft:2016cpw,*Hooft2016_,Betzios_2016,tHooft2018,Hooft:2018gtw,*Hooft:2018syc,*hooft2019,Gaddam_2022}, the interferometer configurations and space-time fluctuation modes proposed here are distinct from those expected from models that use local effective field theory to map the quantum uncertainties in space-time onto definite metric fluctuations\,\cite{Verlinde_2021,Verlinde:2019ade,Banks_2021,Verlinde:2022hhs,banks2023fluc,Zurek_2022,Zurek:2022xzl,LiPRD}.

In recent years, there has been increasing awareness that there are aspects of quantum mechanics and quantum information that simply cannot be captured within a framework based on local field theory, and that a solution for quantum gravity should start with delocalized quantum states and Hilbert spaces rather than fields and metrics~\cite{Banks2011,Banks:2018aed,Hollands_2004,Banks_2020,Banks:2020dus,*banks2022old,banks2023hilbert,*banks2023impossibility,*BanksGRF2023,Giddings_2019_,*Giddings2020,*Giddings2022,Nomura:2012cx,Cao:2017hrv}. This is vividly exemplified through black hole models or various views of cosmic evolution where quantum nonlocality or coherence, entanglement entropy, or the holographic infrared problem has to be considered at the scale of the entire horizon~\cite{Banks:2018jqo,banks2023holography,Giddings_2004,*Giddings_2019,Hooft:2016itl,*Hooft:2016cpw,*Hooft2016_,Betzios_2016,tHooft2018,Hooft:2018gtw,*Hooft:2018syc,*hooft2019,Gaddam_2022,CohenKaplanNelson1999,Marolf:2003bb,hogan2023causal,*hogan2024cosmic,Hogan_2023,Hogan2018a,Hogan_2020,Hogan2022_1,Danielson_2022}. In particular, several colleagues have pointed out that this kind of physics does not just apply to black hole causal horizons or null boundaries; the same holistic aspects of the theory should also apply to flat space-time, especially with careful treatment of finite causal volumes and null surfaces~\cite{Hollands_2004,Banks2011,Banks:2018aed,Banks:2018jqo,banks2023holography,Hollands_2004,Banks_2020,Banks:2020dus,*banks2022old,banks2023hilbert,*banks2023impossibility,*BanksGRF2023,Verlinde_2021,Verlinde:2019ade,Banks_2021,Verlinde:2022hhs,banks2023fluc,krishogan,mackewicz2024,Giddings_2019_,*Giddings2020,*Giddings2022,Kwon:2014,Hogan_2017,Hogan:2015b,Hogan:2016,Danielson_2023,VISSER_2003,Wondrak_2023}. Such arguments resulted in predictions of macroscopic quantum correlations for holographic causal diamonds in flat space-time~\cite{Kwon:2014,Hogan_2017,Hogan:2015b,Hogan:2016,Verlinde_2021,Banks_2021,Verlinde:2022hhs,banks2023fluc}. The earliest proposals by Hogan were founded on conceptual insights and met with skepticism from the community, but the later results are solidly grounded in mainstream, well-established theoretical techniques such as entanglement entropy and bulk-boundary relationships in conformal field theories. As an aside, some of the same characteristic physics was also found in a model that used localized graviton fields (not holographic in their degrees of freedom) but with other exotic ideas from quantum information research such as squeezed vacuum states\,\cite{Parikh:2020nrd,*Parikh_2021,Parikh:2020fhy,Parikh:2023zat,Cho:2021gvg,Kanno:2020usf,*Kanno:2021gpt}. It is promising that many different approaches\,--- heuristic arguments counting holographic degrees of freedom\,\cite{Kwon:2014,Hogan_2017,Hogan:2015b,Hogan:2016}, coherent gravitational shockwaves from quantum mass-energy\,\cite{krishogan,mackewicz2024,Verlinde:2022hhs}, topological mappings between black holes and flat space-time\,\cite{Verlinde_2021}, conformal field theory and entanglement entropy\,\cite{Verlinde:2019ade,Banks_2021,banks2023fluc}, and squeezed graviton states\,\cite{Parikh:2020nrd,*Parikh_2021,Parikh:2020fhy,Parikh:2023zat,Cho:2021gvg,Kanno:2020usf,*Kanno:2021gpt}\,--- have all led to phenomenological regimes similarly reachable by state-of-the-art interferometers. This provides strong motivation for developing more advanced experimental techniques~\cite{GQuESTPublic,McCuller:2022hum,vermeulen2024,grote2020,QIPublic,2013PhRvL.110u3601R,Pradyumna_2020}. The community's recognition of late of this promising target phenomenological regime has spurred much activity, including the funding of a new theory collaboration\,\cite{QuRIOSPublic} and two new experiments being advanced\,\cite{GQuESTPublic,McCuller:2022hum,vermeulen2024,grote2020,QIPublic}. There is also the possibility that solving the nature of quantum space-time and quantum information on causal horizons is the key to resolving the intractable paradoxes we encounter in quantum foundations~\cite{Durham}, or the \:\!$\sim\:\!\!10^{122}$ discrepancy between the value of the cosmological constant and predictions from local QFT\,\cite{Hogan:2020aow,mackewicz2023gravity}. Perhaps most importantly, in the rare observational data where we expect to see active gravitational effects from quantum states, the cosmic microwave background, we already see compelling ``anomalies''\,--- or signatures\,--- that match the features found in our model: exact symmetry of zero correlation at $90^\circ$ and negative correlations at angular separations approaching the antipodes\,\cite{Muir_2018,jones2023universe,Copi_2009,*Schwarz_2016,Hagimoto_2020}, just as one would expect from gravitational back reaction effects if quantum states were nonlocally extended across the horizon. An ansatz designed to match the CMB correlation spectrum at large angular scales was found to map closely onto the angular spectrum predicted here for flat space-time once the inflationary history is taken out from the model, as described in follow-up work\,\cite{hogan2023angular}.\hspace{4pt}

The viability of this proposed experimental design\,\cite{grote2020,QIPublic} makes it a key prong of a strongly motivated multimodal research program, combining foundational theory and phenomenology, reframed statistical analyses of cosmological observations, and a laboratory experiment. We may finally be approaching empirical studies of quantum space-time.\vspace{-3pt}

\section*{\NoCaseChange{Acknowledgments}\vspace{-1pt}}

The author thanks Craig J. Hogan, Stephan S. Meyer, Nathaniel Selub, Lorenzo Aiello, Katherine L. Dooley, Aldo Ejlli, Hartmut Grote, Robert H. Hadfield, Keiko Kokeyama, Denis Martynov, Sander M. Vermeulen, Sougato Bose, Andrew A. Geraci, Gavin W. Morley, and Karim P. Y. Th\'ebault for many helpful discussions. The author is especially grateful to Hartmut Grote for leading the experimental side of this research program.\vspace{-.5pt}

\bibliography{kwonbib} 

\section*{Appendix\vspace{.5pt}}
\subsection*{Relation to Different Approaches to Holography and AdS/CFT\vspace{-1.5pt}}

Discussions of holography in the literature have mostly been dominated by the framework of AdS/CFT, where a precise mathematical description of the duality is possible. The approach in this paper takes the view that the AdS/CFT framework is by construction unable to handle the fundamental issues in flat space-time holography such as the Cohen-Kaplan-Nelson infrared catastrophe arising from the QFT vacuum having too many degrees of freedom in a finite volume of flat space-time~\cite{CohenKaplanNelson1999,banks2023holography,Banks:2018jqo,Banks:2018aed,Banks:2020dus,*banks2022old,banks2023hilbert,*banks2023impossibility,*BanksGRF2023}. In AdS/CFT, the holographic projection is done onto the built-in boundary of AdS space, and as such, the density of states is controlled by the AdS scale, and locality is defined relative to the AdS scale. Taking the asymptotically flat space-time limit and taking the AdS radius to infinity (e.g. ``celestial holography'') would not solve the issue of a Planck resolution background space-time violating the CKN bound at finite scales\,--- according to CKN, the number of states allowed in the QFT scales as \:\!$\sim\:\!\!L^{3/2}$, where $L$ is the size of the space-time volume (even more stringent than the black hole entropy bound). In the formulation of holography adopted here, the role of the AdS scale is instead taken by the causal boundary scale in a finite volume of space-time, and the holographic projection is universally done onto such causal horizons\,--- whether a black hole horizon or a causal diamond in flat space-time (a conformal Killing horizon). For more in-depth discussions of how the built-in AdS scale subtly but fundamentally changes the physical character of the theory, see Refs.\,\cite{banks2023holography,Banks:2020dus,*banks2022old,banks2023hilbert,*banks2023impossibility,*BanksGRF2023}.\vspace{-5.pt}

\subsection*{Implications of Experimental Signatures\vspace{-1.5pt}}

The proposed experiment tests a particular approach to holography, but should not be seen as a general test of holography, especially more mainstream formulations based on AdS/CFT. A positive result (the detection of new physics in the experiment) should be considered strong indication that the holographic principle applies to finite causal diamonds in flat space-time and that the states of holographic quantum space-time are quantized on those causal boundaries with corresponding large coherence scales. It should furthermore be considered a result in support of the Banks-Zurek conjecture\,\cite{Banks_2021} that the fluctuations in the modular Hamiltonian of a causal diamond are equal to the entanglement entropy, which draws from ideas in AdS/CFT but applies them to a broad class of other contexts that include a causal diamond in flat space-time and explicitly excludes a large finite causal diamond in the bulk of AdS space. As such, a positive result should also be considered as (in a sense) disfavoring the traditional formulation of AdS/CFT, in that the holographic projection must be done using the finite causal diamond of the measurement in flat space-time and not in some asymptotically flat limit of AdS space.

The research program is designed to scan a wide range of geometries that can test a broad theory space spanned by different underlying foundational principles~\,\cite{grote2020}. In the case of a detected signature, the correlations in the signal can be modulated by reconfiguring the layout of the detector, which allows us to potentially reverse-engineer the appropriate microstates of a theory of quantum gravity. If a negative result is obtained after an extensive search, this would be considered an invalidation of this particular approach to holography on finite causal boundaries in flat space-time and its corresponding class of theoretical models.\footnote[8]{Certain models by Zurek and colleagues using effective field theory to map quantum space-time operators onto definite metric fluctuations, e.g. the ``pixellon'' model\,\cite{Zurek_2022,Lee_2024,LiPRD}, have already been tested to a normalization prefactor of about $\alpha\lesssim0.1$ using LIGO data around \:\!$\sim\:\!\!40\,$kHz\,\cite{VermeulenLIGO}. However, this is only possible because such models do not capture the true quantum nature of the background space-time: Because the theorized correlations are described by local field theoretic couplings and exclude any modeling of the state reduction that occurs in a measurement of quantum space-time, such models of definite metric fluctuations lead to detector responses that are identical to those for gravitational waves, and therefore can be ruled out with high significance in the same way LIGO attains a high amplification for a classical signal. An exclusion of these limited models should not be considered relevant or applicable to more general models of holographic space-time fluctuations that are founded on more fully quantum treatments of the background space-time, including certain Verlinde-Zurek models based on gravitational shockwaves and the model described in the present manuscript.} However, it would certainly not rule out other approaches to holography that use different boundaries in the implementations, including those originating from AdS/CFT.\vspace{-5.pt}

\enlargethispage{6.5pt} 
\subsection*{Relation to Recent Works on Black Hole Information\vspace{-1.5pt}}

There have been recent works that attempt to resolve the black hole information ``paradox'' using computational complexity arguments~\cite{Almheiri_2019,*Penington:2019npb,Akers:2022qdl,Raju:2020smc}. Using the framework of QFT in curved space-time in the context of AdS/CFT, these works argue that the interior information is hidden (to a semiclassical observer) via scrambling that takes longer than the black hole lifetime to unscramble. The present manuscript takes the view that such a semiclassical framework is foundationally flawed\,\cite{Hollands_2004}, and adopts an alternative description of black hole information by 't~Hooft which combines classical quantum mechanics with general relativity and replaces scrambling with decoherence. Instead of a ``firewall'' that needs to be avoided, the horizon is simply a quantum-classical boundary. In this picture, the cost of restoring unitarity (in a steady-state solution, without invoking the lifetime of the black hole) is that due to the microstates being coherent states on the horizon, to an outside observer the interior might as well not exist. But this is entirely a matter of observer-dependence and what information is accessible or harvested by the observer. 

While there are opinions that these complexity arguments may extend to asymptotically flat space-time, this paper takes the view that approaching this asymptote from the AdS side is fundamentally different from trying to implement holography in a universe with a small positive cosmological constant\,\cite{banks2023holography,Banks:2020dus,*banks2022old,banks2023hilbert,*banks2023impossibility,*BanksGRF2023}. Attempting to resolve the black hole information paradox using AdS/CFT may suffer from inherent limitations by construction. For example, Martinec points out that the proposed resolution using such holographic mapping is extremely nonlocal, highly dependent on the AdS geometry~\cite{martinec2022trouble}. In the alternative picture preferred by this manuscript, the nonlocality on the AdS boundary scale is replaced with nonlocality on a finite causal boundary scale, either black hole horizons or causal diamonds in flat space-time. This is why the Banks-Zurek result on thermodynamic fluctuations of causal horizons\,\cite{Banks_2021} specifically does not apply to black holes in the bulk of AdS space, despite using ideas that originate from AdS/CFT.

As pointed out by Wallace, there is a distinction between the version of the black hole information paradox that concerns the failure of unitarity for a completely evaporating black hole\,\cite{Belot1999} versus the version that concerns the clash in unitarity between the fully statistical-mechanical model of the black hole horizon and the non-unitary description given by QFT, which arises long before complete evaporation~\cite{Wallace_2020}. The model put forth in this work, if successfully tested, would mostly shed light on the latter. The 't~Hooft model adopted here addresses the issue in a way reminiscent of black hole complementarity, but in a more radical form that uses classical quantum mechanics instead of QFT. Our approach agrees with the perspective that the ``firewall'' paradox really arises from the incompatibility between the causal structures of entanglement and space-time, and that a fundamental ingredient for the paradox is the implicit assumption of space-time distinctness\,\cite{Cinti_2021}\,--- that spacelike separated systems are mutually commuting, equivalent to the claim that the definite causal structure of general relativity is imported to a future theory of quantum gravity. A resolution of the paradox forces us to account for connections that are nonlocal for semiclassical physics, and instead of searching for that nonlocality via AdS black holes, we adopt the view that there is nonlocality on the horizon scale, and that information-harvesting by these causal boundaries is associated with quantum-classical transitions.\vspace{-5.pt} 

\enlargethispage{6pt} 
\subsection*{Convergence of Theoretical Frameworks\vspace{-1.5pt}}

In the absence of any known consistent theory of quantum gravity, the research program presented here attempts to conjecture a candidate picture in which ideas and hints from many different streams of physics might come together. Establishing the compatibility of these ideas with one another is an ambitious undertaking that is beyond the scope of this phenomenological work\,--- see e.g, Ref.\,\cite{banks2023hilbert,*banks2023impossibility,*BanksGRF2023} for Banks' attempt to unify insights from the CKN bound, the HST framework, the 't~Hooft black hole algebra, and the recent Banks-Zurek lessons in entanglement entropy and CFT along with previous works from Carlip\,\cite{Carlip_1999}, Solodukhin\,\cite{Solodukhin_1999}, Casini-Huerta-Myers\,\cite{Casini_2011}, and Jacobson-Visser\,\cite{Jacobson_2019}.

It is compelling, however, to note that these different directions of approach can lead to shared (phenomenological) features of what might become an eventual consistent theory of quantum gravity. The idea that quantum gravity has coherent correlations on the scales of casual diamonds in flat space-time can be found in the Banks-Fischler HST framework which assigns Hilbert spaces to causal diamonds to construct a holographic space-time (where the states undergo ``fast scrambling'' at the causal boundary), or in the Verlinde-Zurek picture that uses topological black hole coordinates to connect flat space-time to the 't~Hooft black hole algebra (which describes microstates for which the horizons are boundaries of (de)coherence). The form of the correlations, expanded in spherical harmonics, can be derived from this flat space-time mapping of the 't~Hooft black hole states, a calculation using AdS/CFT, an effective field theory model of the ``pixellon,'' and gravitational shock waves from vacuum fluctuations. The most central result, the random walk scaling in thermodynamic fluctuations of causal diamonds in flat space-time, is shared by several general calculations using entanglement entropy and CFT, the aforementioned models of the pixellon and shock waves from vacuum states, and even models of J-T gravity\,\cite{Gukov_2023} and the fluid-gravity correspondence\,\cite{Bak_2024}. This convergence of different theoretical frameworks makes a powerful case that this set of ideas is worth pursuing as a candidate for a consistent description of quantum gravity, with the remaining pieces of the puzzle to be provided by empirical data.\hspace{6.5pt}

Furthermore, these different models and theories appear to be intriguingly complementary. For example, the 't~Hooft model rigorously captures the quantum mechanical correlations across the horizon of a black hole but struggles to implement its holographic statistical mechanics. This can be addressed by the Banks-Zurek derivation using entanglement entropy and conformal field theory, or the Verlinde-Zurek model using gravitational shockwaves to reproduce the same thermodynamic fluctuations, but neither is able to properly model the physical characteristics of the quantum space-time correlations. While the  't~Hooft algebra helps us more accurately infer the spacelike correlations, the time domain correlations are difficult to calculate from either the 't~Hooft or Verlinde-Zurek models, and the Banks-Fischler HST theory is a perfect complement to the Banks-Zurek model to provide this aspect of the phenomenology. While it may not seem possible to merge these consistently right now, those challenges arise from nothing other than the foundational incompatibilities of their underlying frameworks\,--- QM, QFT, and GR. Nevertheless, it is the widespread perspective of the community that these frameworks must somehow be compatible with one another in the context of a future theory of quantum gravity. The point of this work is to present an experimental research program to interrogate this foundational confluence, by surveying a phenomenology in which several different well-motivated theories from different frameworks give rise to complementary principles and features that may be connected to one another, and describing how observational signatures might be tied to different foundational principles.

\end{document}